\documentclass[12pt,a4paper,tightenlines,nofootinbib]{revtex4}
\usepackage{amssymb}
\usepackage{amsmath}
\usepackage{epsf}
\usepackage{bm}

\newcommand {\si}{\sigma}
\newcommand {\ga}{\gamma}

\newcommand {\de}{\delta}

\newcommand {\al}{\alpha}
\newcommand {\be}{\beta}
\newcommand {\pa}{\partial}

\newcommand {\fr}{\frac}
\newcommand {\W}{\hat\Omega}

\newcommand {\z}{\zeta}

\newcommand {\cf}{{\cal F}}

\newcommand {\pe}{\perp}
\newcommand {\lan}{\langle}
\newcommand {\ran}{\rangle}

\newcommand {\bv}{\mathbf{v}}

\newcommand {\hC}{{C}}

\newcommand {\baga}{\bar{\gamma}}
\newcommand {\bav}{\bar{v}}
\newcommand {\bap}{\bar{p}}
\newcommand {\baw}{\bar{w}}
\newcommand {\baT}{\bar{T}}

\newcommand {\baL}{\bar{L}}

\newcommand {\beg}{\begin{equation}}
\newcommand {\en}{\end{equation}}
\newcommand {\bega}{\begin{eqnarray}}
\newcommand {\ena}{\end{eqnarray}}

\begin{document}

\author{Ariel M\'{e}gevand}
\altaffiliation{Member of CONICET, Argentina}
\email[]{megevand@mdp.edu.ar}
\author{Federico Agust\'{\i}n Membiela}
\altaffiliation{Fellow of CONICET, Argentina}
\email[]{membiela@mdp.edu.ar} \affiliation{IFIMAR (CONICET-UNMdP)}
\affiliation{Departamento de F\'{\i}sica, Facultad de Ciencias
Exactas y Naturales,
  UNMdP, De\'{a}n Funes 3350, (7600) Mar del Plata, Argentina}

\title{Stability of cosmological deflagration fronts}

\begin{abstract}
In a cosmological first-order phase transition,  bubbles of the stable
phase nucleate and expand in the supercooled metastable phase. In many
cases, the growth of bubbles reaches a stationary state, with bubble walls
propagating as detonations or deflagrations. However, these hydrodynamical
solutions may be unstable under corrugation of the interface. Such
instability may drastically alter some of the cosmological consequences of
the phase transition. Here, we study the hydrodynamical stability of
deflagration fronts. We improve upon previous studies by making a more
careful and detailed analysis. In particular, we take into account the
fact that the equation of motion for the phase interface depends
separately on the temperature and fluid velocity on each side of the wall.
Fluid variables on each side of the wall are similar for weakly
first-order phase transitions, but differ significantly for stronger phase
transitions. As a consequence, we find that, for large enough
supercooling, any subsonic wall velocity becomes unstable. Moreover, as
the velocity approaches the speed of sound, perturbations become unstable
on all wavelengths. For smaller supercooling and small wall velocities,
our results agree with those of previous works. Essentially, perturbations
on large wavelengths are unstable, unless the wall velocity is higher than
a critical value. We also find a previously unobserved range of marginally
unstable wavelengths. We analyze the dynamical relevance of the
instabilities, and we estimate the characteristic time and length scales
associated to their growth. We discuss the implications for the
electroweak phase transition and its cosmological consequences.
\end{abstract}

\maketitle

\section{Introduction}

The development of a cosmological first-order phase transition via
nucleation and expansion of bubbles provides an interesting scenario for the
formation of cosmological objects such as magnetic fields \cite{gr01},
topological defects \cite{vs94}, baryon inhomogeneities \cite{w84,h95,ma05},
gravitational waves \cite{gw,lms12}, or the baryon asymmetry of the universe
\cite{ckn93}. One of the relevant aspects of the dynamics of a first-order
phase transition is the motion of the transition fronts. In most cases,
stationary solutions exist, and the bubble growth reaches a terminal
velocity shortly after the bubble nucleates. The velocity of bubble walls
depends on the friction with the plasma \cite{dlhll,micro} and on
hydrodynamics \cite{hidro}. Thus, for a given set of parameters, there can
be one or more solutions. For instance, the wall may propagate as a
supersonic detonation or as a subsonic deflagration. The cosmological
consequences of a phase transition depend strongly on the wall velocity. For
example, detonations favor the generation of gravitational waves, whereas
weak deflagrations favor electroweak baryogenesis. It is well known that
these hydrodynamic solutions may be unstable
\cite{landau,link,hkllm,abney,r96}. Such instabilities would have important
implications for the cosmological remnants of the phase transition.

The standard approach to the stability of combustion or phase
transition fronts is to consider small perturbations of the wall
surface and the fluid \cite{landau}. In the cosmological context, the
stability of deflagrations was first studied in Ref. \cite{link}, in
the non-relativistic limit. This analysis was improved in Ref.
\cite{hkllm} by considering relativistic velocities, and by taking
into account the dependence of the velocity of phase transition
fronts on temperature. The latter is the most important difference
with previous analysis, since temperature fluctuations cause velocity
fluctuations which may stabilize the wall. Numerical simulations
\cite{fa03} agree with this stabilization.

A very simple expression for the wall velocity was considered in Ref.
\cite{hkllm}, namely, $v_w\propto T_c-T_+$, where $T_c$ is the critical
temperature and $T_+$ is the temperature outside the bubble. Perturbing this
equation gives an equation for the perturbations of the interface, which
involves the temperature fluctuations $\delta T_+$. We wish to point out,
however, that such a simple form of the wall velocity does not take into
account (among other things) the dependence on temperature perturbations
$\delta T_-$ behind the wall (which were otherwise considered in
\cite{hkllm} as independent from $\delta T_+$). Furthermore, the results
were applied to a specific case (the electroweak phase transition) without
taking into account the fact that the temperature $T_+$ is higher than the
nucleation temperature $T_N$, due to reheating in front of the wall.

These issues are not relevant in the case of small wall velocity, small
latent heat, and little supercooling, but can be important otherwise. The
discussion on  the electroweak phase transition in Ref. \cite{hkllm} was
carried out for the minimal standard model (SM) with a Higgs mass
$m_H=40GeV$. This gives a rather weak phase transition (although strong
enough for baryogenesis), for which the aforementioned approximations are
valid. The deflagration was found to be stable for wall velocities above a
critical value $v_c=0.07$. Thus, arguing that the wall velocity is generally
$v_w\gtrsim 0.1$, it was concluded that the deflagration is commonly stable.
However, the wall velocity depends on the details of the specific model, and
may be smaller. Furthermore, many extensions of the SM give stronger phase
transitions, which would give higher values of $v_c$.

In this paper we aim to perform a more complete calculation, and to
investigate a wider range of parameters. The main improvements of our
treatment will be to consider a more realistic equation for the wall
velocity, which depends on the fluid fluctuations on both sides of the
interface, and to take into account the effects of reheating, i.e., the fact
that the temperature $T_+$ in front of the wall is not a boundary condition
for the deflagration. In the limit of small wall velocities, little
supercooling, and small latent heat, our results agree with those of Ref.
\cite{hkllm}. Increasing the latent heat or the wall velocity does not
change the qualitative picture. However, as the amount of supercooling is
increased, the critical velocity increases much more quickly than what found
in Ref. \cite{hkllm}, even though the reheating effect tends to stabilize
the deflagration. As a consequence, for strong enough supercooling the wall
propagation is unstable for any velocity below the speed of sound. moreover,
the instability is stronger at the speed of sound. This is in contradiction
with the results of Ref. \cite{hkllm}. The origin of the discrepancy is in
the fact that, in our treatment, the wall velocity depends on fluid
fluctuations on both sides of the wall.

The paper is organized as follows. In the next section we review the
stationary motion of a phase transition front. In Sec. \ref{staban} we
consider linear perturbations of the interface and the fluid. We discuss the
approaches and results of previous works, and then we derive the equations
for the perturbations and find the general solution. In Sec. \ref{stab} we
study the possible instabilities of the deflagration and compare our results
with previous works. We find analytical approximations for the case of small
velocity, small latent heat, and little supercooling. We also discuss how
the reheating which occurs in front of the wall affects the stability of a
deflagration. In Sec. \ref{numres} we use the bag equation of state to study
the instability as a function of the relevant parameters. We explore
numerically a wide region of parameter space. In Sec. \ref{conseq} we
consider the dynamics of the instabilities in a cosmological phase
transition, and in Sec. \ref{model} we discuss the results for the specific
case of the electroweak phase transition. We also discuss briefly on some
cosmological effects. Finally, in Sec. \ref{conclu} we summarize our
conclusions.

\section{Phase transition dynamics and stationary wall motion} \label{stationary}

Cosmological phase transitions are generally a consequence of the
high temperature behaviour of a theory with spontaneous symmetry
breaking. Macroscopically, the system can be described by a
relativistic fluid and a scalar field $\phi$ which acts as an order
parameter. The free energy density $\mathcal{F}(\phi,T)$ has
different minima $\phi_+$ and $\phi_-$ at high and low temperatures,
respectively. These minima characterize two different phases. For
instance, in the case of the electroweak phase transition, $\phi$
corresponds to the expectation value of the Higgs field, and we have
$\phi_+=0$, $\phi_-\sim T_c\sim 100\mathrm{GeV}$.

If the phase transition is first-order, there is a range of
temperatures at which these two minima coexist separated by a
barrier. Thus, the metastable phase is characterized by the free
energy density $ \mathcal{F}_{+}(T)= \mathcal{F}(\phi_+,T)$, whereas
the stable phase is characterized by $\mathcal{F}_{-}(T)=
\mathcal{F}(\phi _{-},T)$. The pressure in each phase is given by
$p_\pm=-\mathcal{F}_\pm$, the entropy density by $s_\pm=dp_\pm/dT$,
and the energy density by $e_\pm=Ts_\pm-p_\pm$. The critical
temperature $T_c$ is defined by
$\mathcal{F}_{+}(T_{c})=\mathcal{F}_{-}(T_{c})$. The latent heat is
defined as the energy discontinuity at $T=T_c$, and is given by $L
=T_c[\mathcal{F}_-'(T_c)-\mathcal{F}_+'(T_c)]$. A first-order phase
transition is characterized by the supercooling of the system (which
remains in the metastable phase below $T_c$), followed by the
nucleation and growth of bubbles of the stable phase at a temperature
$T_N<T_c$ (see, e.g., \cite{gw81,ah92,m00}). The latent heat is
released at the phase transition fronts, which are the bubble walls.

We are interested in the motion of the latter. Therefore, we shall
consider the hydrodynamics of two phases separated by a moving
interface. The equations for the wall and the fluid variables can be
derived from the conservation of the stress tensor for the scalar
field and the fluid. These can be written in the form
  \bega
      \pa_\mu\left(-T\fr{\pa\cf}{\pa T}u^\mu u^\nu+g^{\mu\nu}
      \cf\right)+\pa_\mu\pa^\mu\phi\pa^\nu\phi=0\label{EM1.1},\\
      \pa_\mu\pa^\mu\phi+\fr{\pa\cf}{\pa\phi}+\tilde\eta T_c
      u^\mu\pa_\mu\phi=0\label{EM1.2},
  \ena
with $u^\mu=(\ga,\ga\bv)$ the four velocity of the fluid and $g^{\mu\nu}$
the Minkowsky metric tensor. The terms in parenthesis in Eq. (\ref{EM1.1})
give the well known stress tensor of a relativistic fluid
 \bega
  T^{\mu\nu}=wu^\mu u^\nu-p g^{\mu\nu}, \label{Tmunu}
 \ena
where $w$ is the enthalpy and $p$ the pressure. The last term in Eq.
(\ref{EM1.1}) gives the transfer of energy between the plasma and the field.
The motion of the latter is governed by Eq. (\ref{EM1.2}). The last term in
this equation is a phenomenological damping term\footnote{Recently
\cite{ekns10,hs13,ariel13}, different forms of the damping term have been
proposed in order to account for the saturation of the friction force at
ultra-relativistic velocities \cite{bm09}. Since we shall only deal with
deflagrations, the damping term in Eq. (\ref{EM1.2}) is a good approximation
\cite{ariel13,ms12}.}. The dimensionless coefficient $\tilde\eta$ can be
obtained from microphysics calculations (in the general case, $\tilde\eta$
may depend on the field $\phi$).

For the macroscopic treatment, it is a good approximation to consider
an infinitely thin interface. The system of equations (\ref{EM1.1})
then gives the fluid equations on either of the phases (where the
field is constant), as well as the connection between the solutions
at each side of the interface. On the other hand, Eq. (\ref{EM1.2})
gives an equation for the interface itself and the forces acting on
it. Due to the friction with the surrounding plasma, the bubble walls
in general reach a terminal velocity\footnote{In Ref. \cite{bm09}, it
was shown that, if the wall reaches ultra-relativistic velocities, it
may enter a stage of continuous acceleration. However, models which
allow such ultra-relativistic velocities will not allow, in general,
deflagrations. We are not interested in such models.}. We shall now
consider the stationary motion. In the next section we shall study
perturbations of the stationary solutions. For simplicity, we shall
assume a planar interface moving towards the positive $z$ axis.

\subsection{Fluid equations} \label{hydrostat}

For planar symmetry, the problem becomes $(1+1)$-dimensional, and we
need only consider the $z$ component of the fluid velocity, which we
shall denote $v(z,t)$. Within each phase the field is a constant, and
Eq. (\ref{EM1.1}) just gives the conservation of energy and momentum
of the fluid, $\pa_\mu T^{\mu\nu}=0$, with $T^{\mu\nu}$ given by Eq.
(\ref{Tmunu}). The absence of a distance scale in these equations
justifies to assume the \emph{similarity condition}, namely, that
quantities depend only on the variable $\xi=z/t$. Thus, we have
$\partial _t=-(\xi/t)d/d\xi$ and $\partial _z=(1/t)d/d\xi$.
Furthermore,  variations of thermodynamical quantities are related by
the speed of sound $c_s^2=dp/de$. We have, e.g.,
 \beg
 dp=dw/(1+c_s^{-2}). \label{pwcs}
 \en
We may thus obtain an equation for $v(\xi)$ which depends only on the
parameter $c_s$ \cite{landau} ($c_s$ depends on the equation of state
(EOS) and will be in general a function of temperature). In the
planar case this equation is very simple (see e.g. \cite{lm11}). The
solutions are either $v=$ constant, or the particular solution
$v_{\mathrm{rar}}(\xi)=(\xi-c_s)/(1-\xi c_s)$. The latter corresponds
to a rarefaction wave. In this paper we shall be interested in the
constant velocity solutions.  For these, the temperature is also a
constant.

\subsection{Matching conditions}

\begin{figure}[bth]
\centering
\epsfysize=2cm \leavevmode \epsfbox{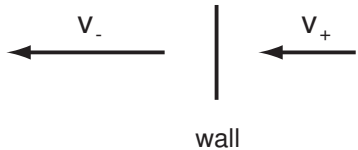}
\caption{Sketch of a deflagration in the wall frame.}
\label{figwall}
\end{figure}

The fluid solutions on each side of the bubble wall can be linked by
integrating Eq. (\ref{EM1.1}) across the wall. It is convenient to
consider a reference frame moving with the wall, where all time
derivatives vanish. We obtain two equations, $\partial_z T^{z0}=0$,
$\partial_zT^{zz}=0$, and the integration gives simply
 \bega
    w_-v_-\ga^2_-&=&w_+v_+\ga^2_+,\label{EM3.1}\\
    w_-v_-^2\ga^2_- +p_-&=&w_+v^2_+\ga^2_+ +p_+ , \label{EM3.2}
 \ena
where $+$ and $-$ signs refer to variables just in front and just
behind the wall, respectively. Notice that in this frame the fluid
velocity is negative (Fig. \ref{figwall}).

These equations have two branches of solutions, called detonations
and deflagrations. Detonations are characterized by the relation
$|v_+|>|v_-|$, whereas deflagrations are characterized by
$|v_+|<|v_-|$. For detonations, the curve of $|v_+|$ vs $|v_-|$ has a
minimum at the Jouguet point $|v_-|=c_{s-}$, where $|v_+|$ takes a
value $v_J^{\mathrm{det}}>c_{s+}$ (hence, for detonations the
incoming flow is supersonic). For deflagrations, the curve of $|v_+|$
vs $|v_-|$ has a maximum value $v_J^{\mathrm{def}}$ at the Jouguet
point $c_{s-}$. In this case, $v_+$ is subsonic,
$v_J^{\mathrm{def}}<c_{s+}$. Detonations are called weak if $v_-$ is
supersonic as well as $v_+$, and deflagrations are called weak if
$v_-$ is subsonic as well as $v_+$. If one of the velocities is
supersonic and the other one subsonic, then the hydrodynamic process
is called strong.

\subsection{Fluid profiles}

The profiles of the fluid velocity and temperature must be
constructed from the solutions of the fluid equations on each side of
the wall, using the matching conditions at the wall and the boundary
conditions. The latter correspond to vanishing fluid velocity far
behind and far in front of the wall. The value of the temperature
$T_N$ far in front of the wall is also a boundary condition. We shall
now describe briefly the possible fluid profiles. For details, see
e.g. \cite{lm11}.

Let us call $\tilde v _+$ and $\tilde v_-$ the values of the fluid
velocity on each side of the phase discontinuity. For a detonation,
the fluid velocity  $\tilde v _+$  vanishes in front of the wall,
which moves supersonically, i.e., we have $v_w=|v_+|\geq
v_J^{\mathrm{det}}>c_{s+}$. Behind the wall, we have a non-vanishing
velocity  $\tilde v _-$, and the wall is followed by a rarefaction
wave. It turns out that only weak detonations can fulfil the boundary
conditions. Therefore, strong detonations are not possible.

\begin{figure}[hbt]
\centering
\epsfysize=4cm \leavevmode \epsfbox{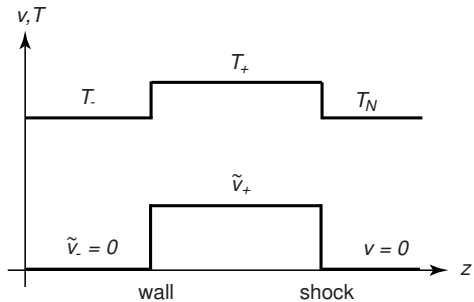}
\caption{Sketch of fluid velocity and temperature profiles for a
deflagration, in the reference frame of the bubble center.}
\label{figdefla}
\end{figure}

For a subsonic wall, the rarefaction solution $v_{\mathrm{rar}}(\xi)$
cannot be accommodated in the velocity profile. The fluid velocity
vanishes behind the wall ($\tilde v _-=0$), and the hydrodynamic
process is a weak deflagration, with $|v_-|=v_w<c_{s-}$. In front of
the wall, the velocity is a constant up to a certain point where the
velocity vanishes abruptly  (see Fig. \ref{figdefla}). Such a
discontinuity without change of phase is called a shock front. At the
shock discontinuity, Eqs. (\ref{EM3.1}-\ref{EM3.2}) still apply, but
now the enthalpy and pressure are related by the same EOS on both
sides of the interface. These equations give the temperature $T_+$ as
a function of the boundary condition $T_N$, as well as the velocity
of the shock front.

Between weak deflagrations and detonations there is a velocity gap
$c_{s-}<v_w< v_J^{\mathrm{det}}$. The ``traditional'' weak
deflagration profile described above is in principle possible for a
supersonic wall as well. In such a case, the hydrodynamic solution is
a strong deflagration, with $|v_-|=v_w>c_{s-}$. Numerical
calculations \cite{ikkl94} suggest that such a strong deflagration is
unstable since, if set as initial condition, it evolves to other
stationary solutions.

\begin{figure}[hbt]
\centering
\epsfysize=3.5cm \leavevmode \epsfbox{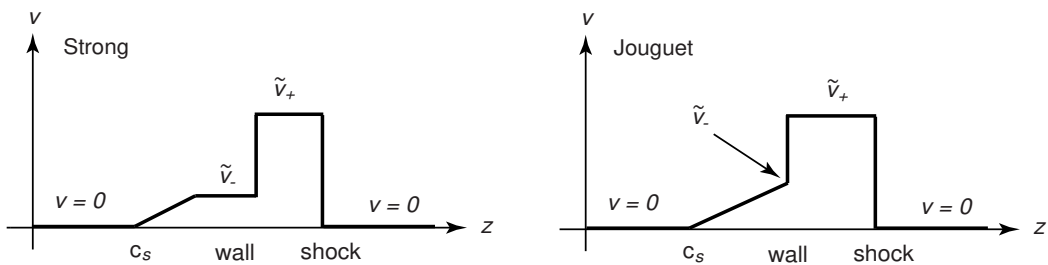}
\caption{Sketch of fluid velocity profile for strong and Jouguet deflagrations.}
\label{figstrong}
\end{figure}

A supersonic deflagration can also be constructed using  the solution
$v_{\mathrm{rar}}$ behind the wall (see Fig. \ref{figstrong}).
Instead of $|v_-|=v_w$, in this case the matching conditions only
require $|v_-|\geq c_{s-}$ (for details, see, e.g., \cite{lm11}).
Therefore, this kind of solution is either a strong or a Jouguet
deflagration. In either case, since the fluid velocity $\tilde{v}_-$
does not vanish, the wall velocity is given by the relativistic sum
of $|v_-|$ and $\tilde{v}_-$, and is always supersonic. In order to
avoid a strong deflagration, the only possibility is a Jouguet
solution, $v_-=-c_{s-}$. As shown in Fig. \ref{figstrong} (right
panel), in this case the rarefaction begins immediately at the wall.
Thus, the strong deflagration shown in the left panel is an
intermediate case between the traditional strong deflagration of Fig.
\ref{figdefla} and the supersonic Jouguet deflagration.

Notice that we have a family of solutions depending on two free
parameters, namely $v_-$ and $\tilde{v}_-$. For $v_-=-v_w$ we obtain
a reduced family of strong deflagrations, namely, those with the
traditional profile, whereas for $v_-=-c_s$ we obtain the family of
supersonic Jouguet deflagrations. Fixing the latter condition and
varying the value of $\tilde{v}_-$, the velocity of the Jouguet
deflagration fills the range of $v_w$ between the weak deflagration
and the detonation.

\subsection{Equation for the interface}

We shall now obtain an equation for the interface from Eq. (\ref{EM1.2}). In
the reference frame of the wall,  we multiply by $\phi'\equiv d\phi/dz$ and
then  integrate across the interface. We obtain
 \beg\label{ecfric}
    p_- - p_+ + \int_-^+ \left(-\fr{\pa\cf}{\pa T}\right)\fr{dT}{dz}dz
   +\tilde\eta T_c \int_-^+ \phi'^2 \ga v dz=0 ,
 \en
where we have used the relation $(\pa \cf / \pa \phi) d\phi/dz=d\cf / dz -
(\pa\cf/\pa T)dT/dz$. In Eq. (\ref{ecfric}) we identify the force which
drives the bubble expansion,
 \beg
   F_{\mathrm{dr}} =  p_-(T_-)-p_+(T_+)+\int_-^+ \frac{\partial\mathcal{F}}{\pa T}
   \frac{d T}{dz}dz , \label{fdrexac}
 \en
and the friction force, which explicitly depends on the velocity of
the fluid with respect to the wall.  Thus, Eq. (\ref{ecfric}) gives
the balance between the forces that rule the stationary wall motion.

While  the integration of Eqs. (\ref{EM1.1}) across the wall was
straightforward, in this case we shall need some approximations to
avoid the dependence on the wall shape \cite{ms09}. To evaluate the
last integral in Eq. (\ref{ecfric}), we notice that $\phi'(z)^2$
behaves approximately like a delta function which picks the value of
$\gamma v$ at the center of the wall width, say $\gamma_0 v_0$. Thus,
the integral gives $\gamma_0v_0\sigma$, where
$\sigma\equiv\int\phi'^2dz$ is the surface tension. We shall
approximate the value of $\gamma_0 v_0$ by the average  $\lan \gamma
v\ran\equiv \fr{1}{2}(\gamma_+v_+ +\gamma_-v_-)$. Thus, the last term
in (\ref{ecfric}) can be written in the form $-\eta \lan \gamma
v\ran$, with
 \beg
    \eta=\tilde\eta T_c\sigma ,
 \en
and the force balance reads
  \beg\label{solfric}
   {\eta} \lan \ga v\ran=-{F_{\mathrm{dr}}}.
  \en

For the first integral in Eq. (\ref{ecfric}), we can use a linear
approximation for the $z$ dependence of the entropy density $s=-\partial
\mathcal{F}/\partial T$ inside the wall. We obtain $\fr{1}{2}(s_+ +s_-)(T_+
-T_-)$. This gives the following approximation for the driving force,
 \beg
 F_{\mathrm{dr}}=p_-(T_-)-p_+(T_+) + \lan {s}\ran(T_+-T_-). \label{fdr}
 \en
In many cases, the free energy has even powers of the temperature, whereas
the last term in Eq. (\ref{fdr}) introduces odd powers. A more convenient
approximation for such cases can be obtained by noticing that $(\partial
\mathcal{F}/\partial T)dT=(\partial \mathcal{F}/\partial T^2)dT^2$. Then,
instead of using a linear approximation for $\partial \mathcal{F}/\partial
T$, we may use a linear approximation for $\partial \mathcal{F}/\partial
T^2$. We obtain
 \beg
F_{\mathrm{dr}}=p_-(T_-)-p_+(T_+)
+\left\langle\fr{dp}{dT^2}\right\rangle\left(T_+^2-T_-^2\right).
\label{fdr2}
 \en
The approximations (\ref{fdr}) and (\ref{fdr2}) are quantitatively very
similar, though analytically different. The former involves the physical
quantity $s$ and is useful for physical discussions. As we shall see, the
latter gives cleaner analytical results.

\section{Stability analysis} \label{staban}

\subsection{Previous results and relevant scales} \label{scales}

The stability of deflagrations in a cosmological phase transition was first
studied by Link \cite{link} in the non-relativistic limit. The results are
similar to those of a classical gas \cite{landau} (with the enthalpy taking
the role of the mass density). According to this analysis, small wavelength
perturbations of the phase transition front are stabilized by the surface
tension $\sigma$, whereas large wavelength perturbations grow exponentially.
The perturbations are of the form
 \beg
  e^{i\mathbf{k}\cdot \mathbf{x}^{\bot}+qz+\Omega t},
 \en
and the initial growth time is given by $\Omega^{-1}$. The essential
features of the instability are easily seen if we consider the limit
$T_\pm\simeq T_c$ in Link's result, so that the latent heat is given by
$L\simeq w_+-w_-$. For small $L/w_+$ we have
 \beg
 \Omega\simeq
 \fr{L}{w_+}\fr{v_w}{2}\left(1-\fr{k}{k_c}\right)k,
 \label{omegalink}
 \en
with
 \beg
  k_c={L}{v_w^2}/{\sigma}
 \en
Thus, the critical wavelength above which perturbations are unstable
is given by  $\lambda_c=1/k_c$. Notice that, for non-relativistic
velocities, Eqs. (\ref{EM3.1}-\ref{EM3.2}) give $p_+-p_-=Lv_+v_-$
which, for small $L/w_+$ becomes $p_+-p_-=Lv_w^2$. Therefore, the
critical wavelength can be written as
 \beg
 \lambda_c=\frac{\sigma}{p_+(T_+)-p_-(T_-)}\equiv d_c. \label{dc}
 \en
Physically, $d_c$ is the length scale over which  the surface tension
just balances the difference of the pressures on each side of the
interface \cite{landau,link}.

This result was improved by Huet et al. \cite{hkllm}. In the first
place, relativistic velocities were considered in order to study fast
deflagrations ($v_w\to c_s$). Most important, the dependence of the
wall velocity on temperature  was taken into account. A very simple
form for the velocity was considered.
 \beg
  \gamma_w v_w(T_+)=\frac{p_-(T_+)-p_+(T_+)}{\eta}
  \approx  \frac{L(1-T_+/T_c)}{\eta}.
  \label{vwsimple}
 \en
where $\eta$ is a friction parameter, which can be assumed to be a
constant. The last approximation in Eq. (\ref{vwsimple}) is valid in
the non-relativistic limit and for $T_+$ close to $T_c$. The
inclusion of Eq. (\ref{vwsimple}) into consideration takes into
account that temperature fluctuations induce a change in the velocity
of the interface. Instead of this, the treatment of Landau
\cite{landau} just equated the normal velocity fluctuations $\delta
v_\pm$ of the fluid  to the velocity of the surface perturbation
$\partial_0 \zeta$, while Link \cite{link} considered the energy flux
$F=w v\gamma ^2$ to be proportional to the net blackbody energy flux,
$F\propto g_-(\pi^2/30) (T_+^4-T_-^4)$, where $g_-$ is the effective
number of degrees of freedom in the $-$ phase. Notice that these
conditions lack important information, namely, that the interface is
a phase transition front (and not, e.g., a burning front). This
information is not present in either of the equations considered in
\cite{landau,link}. In contrast, in Eq. (\ref{vwsimple}), the
velocity of the interface depends on the pressure difference and
vanishes at $T_+=T_c$, which is the essential feature of a phase
transition.

From Eq. (\ref{vwsimple}), an important parameter arises in the
perturbation equations, namely,
 \beg
  \beta= T_c\left(-\frac{dv_w}{dT_+}\right)\gamma_w^2
  v_w. \label{betasimple}
 \en
The dependence of the wall velocity on temperature tends to stabilize the
perturbations. In the non-relativistic limit and for $\Omega\ll k$, the
result of Ref. \cite{hkllm} takes the form
 \beg
 {\Omega}\simeq
 \frac{\baL v_w}{2}\frac{\lambda-\lambda_c}{\lambda^2}\frac{(1-\beta)}{1+\beta \baL+d_b k},
 \label{omegaellosnr}
 \en
with
 \beg
  \lambda_c=d_c\left(1+\frac{2d_b/d_c}{\baL(1-\beta)}\right),
  \label{lambdacnrellos}
 \en
where $\baL=L/w_+$, and a new length scale
 \beg
  d_b=\frac{\sigma}{p_-(T_+)-p_+(T_+)} \label{db}
 \en
has appeared due to the introduction of Eq. (\ref{vwsimple}). Its physical
significance is similar to that of $d_c$ [cf. Eq. (\ref{dc})]. It gives the
length scale over which the  surface tension just balances the pressure
difference \emph{in equilibrium at $T=T_+$}. Thus, it gives the size scale
of a critical bubble before it begins to grow. As remarked in Ref.
\cite{hkllm}, the quantity $d_b$ gives also a time scale characteristic of
the dynamics of growth, corresponding to the acceleration period before the
bubble wall reaches a terminal velocity. Within the present approximations,
the two scales $d_b$ and $d_c$ are related by
 \beg
  \frac{d_b}{d_c}=\frac{v_w^2}{1-T_+/T_c}, \label{dbdc}
 \en
and Eq.  (\ref{betasimple}) gives
 $\beta =v_w L/\eta$.
We can eliminate the friction parameter and write \cite{hkllm}
 \beg
  \beta={v_w^2}/{v_c^2},
 \en
where
 \beg
  v_c^2\equiv(T_c-T_+)/T_c. \label{vc}
 \en
The length scales are thus related by $d_b/d_c=\beta$.

A few comments on these scales are worth. Notice that, in the equilibrium
situation with $T_-=T_+$, we have $p_-(T_+)>p_+(T_+)$, whereas, for a real
deflagration with $T_+>T_-$ (see Fig. \ref{figdefla}), the pressure balance
is inverted, $p_-(T_-)<p_+(T_+)$ (this is why it is necessary to invert the
order of $p_-$ and $p_+$ between the definitions of $d_c$ and $d_b$). It is
evident that the pressure difference $p_-(T_-)-p_+(T_+)$, being negative,
cannot be used as the driving force, and the equilibrium value
$p_-(T_+)-p_+(T_+)$ in Eq. (\ref{vwsimple}) is a better approximation
(although it does not take hydrodynamics into account). A still better
approximation to the driving force is given by Eq. (\ref{fdr}).

In practice, the relevant difference between the results of Refs.
\cite{link} and \cite{hkllm} is the appearance of the quantity $(1-\beta)$
in Eqs. (\ref{omegaellosnr}-\ref{lambdacnrellos}) with respect to Eqs.
(\ref{omegalink}-\ref{dc}). For $\beta\ll 1$, the results are essentially
the same. However, for $\beta\approx 1$ the situation changes drastically.
For $\beta<1$, perturbations on wavelengths $\lambda<\lambda_c$ are stable.
Thus, as $\beta$ approaches 1, only very large wavelengths will be unstable,
and with very large growth time $\Omega^{-1}$. For $\beta>1$, the behavior
is inverted. Perturbations with $\lambda>\lambda_c$ are now {stable}.
Besides, the second term in Eq. (\ref{lambdacnrellos}) dominates, and we
have $\lambda_c<0$. This means that \emph{perturbations at all scales are
stable for $\beta>1$}. Hence, $v_c$ is a critical velocity, above which the
deflagration becomes stable.

Notice that the second term in Eq. (\ref{lambdacnrellos}) is not a small
correction, even for $\beta\ll 1$. In fact, due to the smallness of $\baL$,
this term will generally dominate. As a consequence, the critical wavelength
predicted by Huet et al. is generally quite larger than Link's result. A
numerical study of the stability of planar walls was performed in Ref.
\cite{fa03} for the QCD phase transition, for a value $\baL\approx 0.089$.
For the case of deflagrations, two cases were considered, corresponding to
$v_w=0.196$ (case A) and  $v_w=0.048$ (case B). For each of these two runs
of the numerical simulation, the planar wall was initialized with sinusoidal
perturbations of wavelengths up to $\lambda=5\lambda_c^{\mathrm{Link}}$,
where $\lambda_c^{\mathrm{Link}}$ is the critical wavelength obtained by
Link and given approximately by Eq. (\ref{dc}). According to Link's results,
these perturbations should be unstable. However, the perturbations decayed
in the simulation, in agreement with the results of Huet et al. Indeed, the
stability parameter is $\beta=0.516$ for case A and $\beta=0.722$ for case B
\cite{fa90}. Hence, according to Eq. (\ref{lambdacnrellos}), we have
$\lambda_c\approx 25\lambda_c^{\mathrm{Link}}$  and $\lambda_c\approx
60\lambda_c^{\mathrm{Link}}$, respectively. Perturbations of wavelengths
$\lambda$ higher than these values of $\lambda_c$ should be unstable
according to the results of Huet et al. Unfortunately, such perturbations
were not considered in the simulations of Ref. \cite{fa03}.

Although the treatment of Ref. \cite{hkllm} improved significantly
upon previous stability analysis, some of the approximations used in
this approach will not hold in the general situation. The most
important are the following.

In the first place, we remark that the simple expression
(\ref{vwsimple}) for $v_w$ implicitly assumes the relations $T_-=T_+$
and $v_+=v_-=v_w$ [cf. Eqs. (\ref{solfric}),(\ref{fdr})], while for a
deflagration we have $T_-<T_+$ and $ v_+<v_-=v_w$. Moreover, since
the expression (\ref{vwsimple}) depends only on $T_+$ and $v_-$, it
does not allow to perform \emph{independent perturbations} of
variables on each side of the wall (such as $\delta T_-,\delta v_+$).
These limitations of the surface equation constrain the validity of
the treatment of Huet et al. This contrasts with their treatment of
the fluid equations, where independent perturbations were considered
for $-$ and $+$ variables. One expects that the approximation $\delta
T_-=\delta T_+$ will be valid if $T_-\simeq T_+$. However, the latter
is not the most general case.

In the second place, the reheating in front of the wall was not taken
into account. The value of $T_+$ was estimated  from results on the
amount of supercooling\footnote{The temperature at which bubbles
nucleate and expand can be estimated using the bubble nucleation
rate, which is calculated using the thermal instanton technique
\cite{linde}.} for the electroweak phase transition \cite{dlhll}
(i.e., the approximation $T_+\simeq T_N$ was used). Notice that, for
a deflagration, the fluid is reheated in front of the wall (see Fig.
\ref{figdefla}). As a consequence, the temperature $T_+$ in front of
the phase transition front does not coincide with the nucleation
temperature $T_N$. Depending on the wall velocity and the amount of
latent heat released, the local reheating can be important.

Our derivation of the perturbation equations will be similar to that
of Ref. \cite{hkllm}. The main difference will be, essentially, that
instead of considering Eq. (\ref{vwsimple}), we shall consider Eq.
(\ref{solfric}), which depends explicitly on the two velocities
$v_\pm$ and on the two temperatures $T_\pm$ (through the driving
force). According to the approximation (\ref{fdr}), we have
$F_{\mathrm{dr}}=p_-(T_-)-p_+(T_+) +\langle s\rangle (T_+-T_-)$. In
the case of small supercooling (i.e., $T_c-T_{\pm}\ll T_c$) and small
wall velocity, the pressure difference is $\mathcal{O}(Lv_w^2)$ and
can be neglected in comparison with the term $\langle s\rangle
(T_+-T_-)$. Besides, we can use the approximation \cite{ms09}
 \beg \label{tmatmenr}
 T_+-T_-=\fr{\Delta
s(T_c)}{s_-(T_c)}(T_c-T_+)
 \en
to obtain
 \beg
  F_{\mathrm{dr}}=({\langle w\rangle}/{w_-})\,{L(1-T_+/T_c)}, \label{fdrsmallsup}
 \en
which, taking into account that  $\langle v\rangle=-v_w\langle
w\rangle/w_+$, gives
 \beg
  v_w=\frac{w_+}{w_-}\frac{L(1-T_+/T_c)}{\eta}.\label{vwsmallsup}
 \en
This is similar to Eq. (\ref{vwsimple}), except for the factor
$w_+/w_-$, which is $\approx 1$ for small latent heat. Thus, in the
limit of small supercooling, small $v_w$, and small $L/w_+$, we
obtain the approximation used in Ref. \cite{hkllm} for the wall
velocity. In our treatment, the parameter  $d_b$ will be replaced by
\begin{equation}
d=\frac{\sigma}{F_{\mathrm{dr}}} \label{d}
\end{equation}
For small supercooling and small latent heat, we have $d_b\approx d$.
According to Eq. (\ref{dbdc}), the parameter $d_c$ is determined by $d_b$
and $v_c$. Similarly, our results will depend on $d$ and the critical
velocity.

\subsection{Linearized  equations}

We shall consider small perturbations of the fluid and the interface.
The planar symmetry allows us to consider a single transverse
direction $x^{\perp}$ instead of two directions $x,y$. The
perturbation variables we shall use are the pressure fluctuation $\de
p(x^{\pe},z,t)$, the variation of the velocity along the wall motion
$\de v (x^{\pe},z,t)$, the transverse velocity $v^\pe(x^{\pe},z,t)$,
and the variation of the wall position $\z(x^\pe,t)$. For the sake of
clarity, in this subsection we shall denote the stationary solutions
with a bar. Thus, we have
 \bega\label{defs.1}
  p&=&\bar{p}+\de p, \\
  u^\mu&=&\bar{u}^\mu+\de u^\mu=(\bar{\ga},0,0,\bar{\ga}\bar{v})+
  (\bar{\ga}^3\bar{v}\de v,\bar{\ga}v^\pe,\bar{\ga}\de v), \label{defs.2} \\
  z_w&=&\bar{z}_w +\z. \label{defs.3}
 \ena
To derive the equations for these four variables, we shall consider
again Eqs. (\ref{EM1.1}-\ref{EM1.2}) for a field configuration
corresponding to the perturbed wall.

\subsubsection{Fluid equations}

Away from the (perturbed) interface, the field is a constant as
before, and Eq. (\ref{EM1.1}) gives the local conservation of energy
and momentum $\pa_{\mu}T^{\mu\nu}=0$, where $T^{\mu\nu}$ is given by
Eq. (\ref{Tmunu}). To obtain the fluid equations it is convenient to
take the projections of the 4-divergence of the stress tensor along
the directions of the fluid 4-velocity and orthogonal to it, $
  u_\mu {T^{\mu\nu}}_{,\nu}=0\label{HD2}, \
  (g_{\al\mu}-u_\al u_\mu){T^{\mu\nu}}_{,\nu}=0\label{HD3}.
$
Taking into account the relation between enthalpy and pressure
variations, Eq. (\ref{pwcs}), we obtain
 \bega
  c_s^2 w u^\nu_{,\nu}+u^\nu p_{,\nu}=0,\label{HD4}\\
  w u_{\al,\nu}u^\nu-p_{,\al}+u_\al u^\nu p_{,\nu}=0. \label{HD5}
 \ena
We are interested in the stability of the deflagration solution
depicted in Fig. \ref{figdefla}, for which the fluid profile is
constant on both sides of the wall. Therefore, we must consider
perturbations from a constant solution (we shall not consider here
perturbations of the shock front). To linear order in the
perturbations we obtain, for the direction along the fluid velocity,
 \beg\label{HD6}
    c_s^2 \bar{w}(\bar{\ga}^2\bar{v}\de v{,_0}+\bar{\ga}^2\de v_{,z}+v^\pe_{,\pe})+\de p_{,0}+\bar{v}\de
    p_{,z}=0
 \en
and, for the orthogonal directions,
 \bega
  \bar{w}\bar{\ga} ^2(\de v_{,0}+\bar{v}\de v_{,z})+\bar{v}\de p_{,0}+\de p_{,z}=0, \label{HD7} \\
  \bar{w} \bar{\ga}^2(v^\pe_{,0}+\bar{v}v^\pe_{,z})+\de p_{,\pe}=0. \label{HD8}
 \ena
We have not specified the reference frame yet. We shall consider the
frame which moves with the unperturbed wall. Therefore, $\bav$
corresponds to the incoming and outgoing flow velocities.

\subsubsection{Matching conditions at the interface}

To obtain matching conditions for the perturbations, we need to
derive the matching conditions for the perturbed wall, and then
subtract those for the stationary wall. We shall consider that the
unperturbed wall is at $\bar{z}_w=0$ and the perturbed wall at
$z_w=\z (x^\pe,t)$. Let us consider again Eqs. (\ref{EM1.1}),
$\pa_\mu(wu^{\mu}u^{\nu}-g^{\mu\nu}p)=-\Box\phi\pa^{\nu}\phi$.

Since there is a discontinuity at the bubble wall, if we integrate
these equations  in a small interval along the normal direction
across the surface, only the normal derivatives will give a finite
difference. Thus we can neglect all other derivatives. We obtain
 \beg
   \pa_n(w\ga^2  v^n)=0, \label{stat.1}
\;\;
   \pa_n(w\ga^2  v^n v^{\perp})=0, 
\;\;
   \pa_n(w\ga^2  v^{n2}+p)=\Box \phi \ \pa_n\phi. 
 \en
For instance, for the unperturbed wall we have $n=z$, $\Box
\phi=-\pa_z^2\phi$, and we re-obtain Eqs. (\ref{EM3.1}-\ref{EM3.2}),
 \bega
   \Delta ( \baw\bav\baga^2)&=&0,\label{EM3.1b}\\
   \Delta (\baw\bav^2\baga^2 +\bap)&=&0, \label{EM3.2b}
 \ena
together with the continuity of the transverse velocity, $\Delta
\bav^{\perp}=0$. The velocity $\bav^{\perp}$, though, is set to zero
by symmetry. If we use Eqs. (\ref{stat.1}) for the perturbed wall, we
then have to express $\partial_n$ in terms of $\partial_z$,
$\partial_{\perp}$ and $\partial_0$, in order to compare with the
stationary equations.

Alternatively,  we may stay in the coordinate system of the
unperturbed wall and just consider the fluid equations,
 \bega
   \pa_0 (w \ga^2 - p)+\pa_z(w\ga^2  v)+\pa_{\perp}(w\ga^2  v^{\perp})&=&-\Box \phi \ \pa_0\phi,
     \label{EMcomp.1}\\
   \pa_0 (w \ga^2 v^{\perp})+\pa_z(w\ga^2  v v^{\perp})+\pa_{\perp}(w\ga^2  v^{\perp}+p)&=&\Box \phi \
   \pa_{\perp}\phi,
     \label{EMcomp.2}\\
   \pa_0 (w \ga^2 v)+\pa_z(w\ga^2  v^2+p)+\pa_{\perp}(w\ga^2 v v^{\perp})&=&\Box \phi \
   \pa_z\phi,
     \label{EMcomp.3}
 \ena
taking into account the discontinuity of the variables at the wall.
We shall use this approach. We are going to integrate across the wall
along the $z$ axis. Since we shall integrate in a vanishingly small
interval, we may neglect any dependence on $x^\pe$ and $t$, other
than the position of the discontinuity, i.e., we may assume (for this
integration) step functions depending only on $z-\z (x^\pe,t)$.
Hence, we have, e.g., $\pa_0 w=-\pa_0\z \pa_z w$, $\pa_{\perp}
w=-\pa_{\perp}\z \pa_z w$, etc. This leaves only $z$ derivatives in
the lhs of Eqs. (\ref{EMcomp.1}-\ref{EMcomp.3}).

Keeping up to linear terms in the perturbations $\z$ and $v^{\perp}$,
only terms proportional to $\pa_z^2\phi \ \pa_z\phi$ (which vanish
after integration) remain in the rhs, except in Eq. (\ref{EMcomp.3}),
where there is also a term proportional to $(\pa_z\phi)^2$.
Performing the $z$ integration, the latter gives the surface tension
$\sigma=\int (\pa_z\phi)^2dz$. We thus obtain
 \bega
      -\pa_0\z \Delta(w\ga^2 - p)+\Delta(w \ga^2 v)&=&0,\label{EM2.1}\\
      \Delta(w\ga^2 v  v^\pe)-\pa_\pe\z\Delta p &=&0,\label{EM2.2}\\
      \Delta(w\ga^2 v^2 +p)&=&-\si(\pa_0^2-\pa_\pe^2)\z,\label{EM2.3}
 \ena
where $\Delta$ applied to any function $f$ means $\Delta f=f_+-f_-$.
In the last equation we have used the fact that, according to Eq.
(\ref{EM2.1}), $\Delta(w \ga^2 v)=\mathcal{O}(\z)$. Now we replace
$w=\baw+\delta w$, $v=\bav+\delta v$, etc., taking into account the
unperturbed equations (\ref{EM3.1b}-\ref{EM3.2b}) and the relation
$\delta w=(1+c_s^{-2})\delta p$. To first order in all the
perturbations, we have
 \bega
    \Delta\left[\baw\baga^2(1+\bav^2)(-\pa_0\z+\baga^2\de v)+
    (1+c_s^{-2})\baga^2\bav\de p\right]=0,\label{EM4.1}\\
    \Delta(v^\pe+\bav\pa_\pe\z)=0,\label{EM4.2}\\
    \sigma(\pa_0^2-\pa_\pe^2)\z+\Delta\left[2\baw\baga^4\bav\de v+
    \left(1+(1+c_s^{-2})\baga^2\bav^2\right)\de p\right]=0.
    \label{EM4.3}
 \ena

\subsubsection{Equation for the interface}

Let us now consider the field equation (\ref{EM1.2}). The field
varies in a region (the wall width) around the wall position
$z_w=\zeta(x^{\perp},t)$. Hence, we may assume a field profile of the
form\footnote{Notice that here we are considering a reference frame
for which $\bar{z}_w=\dot{\bar{z}}_w=0$, and $\zeta$ is a small
perturbation, so we have $\gamma_w=1+\mathcal{O}(\zeta^2)$.}
$\phi(z,x^{\perp},t)=\phi[z-\z(x^{\perp},t)]$. To first order in $\z$
and $v^{\pe}$, we have
$\pa_{\mu}\pa^{\mu}\phi=\phi'(\pa_{\pe}^2-\pa_0^2)\z-\phi''$ and
$u^{\mu}\pa_{\mu}\phi=\gamma(-\pa_0\zeta+v)\phi'$. Multiplying  Eq.
(\ref{EM1.2}) by $\phi'(z-\z)$ and integrating in $z$ (as we did in
Sec. \ref{stationary}), we obtain
 \beg\label{EM5}
    \si(\pa_0^2-\pa_\pe^2)\z=    p_- - p_+ + \int sdT+\tilde\eta T_c\int \phi'^2
    \ga(v-\pa_0\z)dz.
 \en
This equation is similar to Eq. (\ref{ecfric}), except for the three
terms depending on $\z$. The first one, $\si\pa_0^2\z$, takes into
account the acceleration of a surface element of the wall which,
according to Eq. (\ref{EM5}), is determined by the sum of all the
forces acting on it. The second one, $-\si\pa_\pe^2\z$, gives the
restoring force due to the curvature of the surface. Finally, the
term $-\pa_0\z$ takes into account the fact that the friction force
depends on the relative velocity $v_r$ between the fluid and the
wall. We have $\gamma(v-\pa_0\z)=\ga_r v_r$.

Approximating the integrals in  (\ref{EM5}) as we did in Sec.
\ref{stationary}, we obtain
  \beg\label{EM7}
  \si(\pa_0^2-\pa_\pe^2)\z- F_{\mathrm{dr}}-\eta \lan
  \ga(v-\pa_0\z)\ran=0.
  \en
The various thermodynamical quantities (entropy, pressure,
temperature), are  related through the equation of state. Hence, we
may consider the driving force as a function of $T_-$ and $T_+$. In
the stationary case, Eq. (\ref{EM7}) gives Eq. (\ref{solfric}),
  \beg\label{EM8}
    \lan \baga\bav\ran=-{F_{\mathrm{dr}}(\bar{T}_+,\bar{T}_-)}/{\eta},
  \en
where $\bar{T}_-$ and $\bar{T}_+$ are related through Eqs.
(\ref{EM3.1b}-\ref{EM3.2b}). For the perturbations we obtain
 \beg\label{EM9}
    \si(\pa_0^2-\pa_\pe^2)\z=
    \eta\lan\de(\ga v)-\baga\pa_0\z\ran +\fr{\pa F_{\mathrm{dr}}}{\pa T_+}\delta T_+
    +\fr{\pa F_{\mathrm{dr}}}{\pa T_-}\delta T_-,
 \en
The derivatives $\partial F_{\mathrm{dr}}/\partial T_\pm$ can be
calculated by using either the approximation (\ref{fdr}) or the
approximation (\ref{fdr2}). In terms of our perturbation variables
$\delta p_\pm$, we have
 \beg
  \fr{\si}{\eta}(\pa_0^2-\pa_\pe^2)\z =
  \fr{1}{2}\left[\de (\gamma_+ v_+) + b_+\fr{\de p_+}{w_{+}} -
  \baga_+\pa_0\z \right]
  +  \fr{1}{2}\left[\de (\gamma_-  v_-) +b_-\fr{\de p_-}{w_-} -
   \baga_-\pa_0\z \right], \label{EM13}
 \en
where
 \beg \label{betapm}
  \fr{b_{\pm}}{2}\equiv  T_{\pm}\fr{1}{\eta}\fr{\pa F_{\mathrm{dr}}}{\pa
  T_{\pm}} .
 \en
The parameters $\sigma$ and $\eta$ can be written in terms of the
fluid velocity and the scale $d$ using Eqs. (\ref{EM8}) and
(\ref{d}). Thus, we can write Eq. (\ref{EM13}) in a more concise
form,
 \beg \label{EM17}
 \left\langle \baga\bav d (\pa_0^2-\pa_\pe^2)\z + \baga^3\de
  v +b\de p/w-   \baga\pa_0\z\right\rangle=0,
 \en
with
 \beg
 \label{betapm2}
  \fr{b_{\pm}}{2}= \lan - \baga\bav\ran \fr{T_{\pm}}{F_{\mathrm{dr}}}
  \fr{\pa F_{\mathrm{dr}}}{\pa
  T_{\pm}} =2\lan - \baga\bav\ran \fr{T^2_{\pm}}{F_{\mathrm{dr}}}
  \fr{\pa F_{\mathrm{dr}}}{\pa
  T^2_{\pm}} .
 \en
The last equality is useful if $F_{\mathrm{dr}}$ is quadratic in
temperature.

To understand the meaning of these equations, it is useful to
consider some simplifications used in previous works. In Ref.
\cite{hkllm} hydrodynamics was neglected in the equation for the
wall. This means that, in the first place, it was assumed that
$\baT_-=\baT_+$ and $\bav_+=\bav_-=-v_w$. In this case we have
$F_{\mathrm{dr}}=p_-(\baT_+)-p_+(\baT_+)$ and the scale $d$ becomes
equal to $d_b$ defined in (\ref{db}). In the second place, only
linearly-dependent perturbations were considered for the interface
equation, $\de T_-=\de T_+, \de v_-=\de v_+$, such that even the
perturbed force is of the form
$F_{\mathrm{dr}}(T_+)=p_-(T_+)-p_+(T_+)$. This is the most sensitive
simplification used in Ref. \cite{hkllm}, since perturbations in each
phase may be quite different\footnote{This is more apparent in the
case of detonations, for which perturbations can grow \emph{only} in
the $-$ phase, and leads the authors of Ref. \cite{hkllm} to a wrong
conclusion about the stability of detonations \cite{abney,
stabdeto}.}. With these approximations, the two terms in the rhs of
(\ref{EM13}) are almost identical. The sum of the two coefficients
$b_{\pm}/2$ contains a \emph{total} derivative
$dF_{\mathrm{dr}}/dT_+$ and gives
 \beg
  \lan b\ran= T_{+}\fr{d (\gamma_w v_w)}{d T_{+}} .
 \en
We thus obtain
 \beg \label{ecbetaellos}
   v_w d_b(\pa_0^2-\pa_\pe^2)\z = \left(1-\beta\right)\gamma_+^2\de
  v_+ -   \pa_0\z,
 \en
where
 \beg \label{betaellos}
  \beta=- T_{+}\fr{\pa v_w}{\pa T_{+}}\fr{1}{w_+}\fr{\de p_+}{\de v_+}.
 \en
In Eq. (\ref{betaellos}), the variation $\de p/\de v$ depends on the
solution of the fluid equations (\ref{HD6}-\ref{HD8}) on the $+$ side of the
wall.

Equation (\ref{ecbetaellos}) is essentially\footnote{There is a
discrepancy, namely, the relative factor of $\gamma_+^2$ between the
terms $\de v_+$ and $\pa_0\z$. The origin of this is that, in their
derivation, the authors of \cite{hkllm} considered, for $v_+\approx
-v_w$, the relation $\de(v_++v_w)=\pa_0\z$. However, the correct
relativistic velocity sum gives $\de(\gamma_+^2(v_++v_w))=\pa_0\z.$}
the same as Eq (52) of Ref. \cite{hkllm}. As we have seen at the
beginning of this section, the coefficient $\beta$ plays a relevant
role in the hydrodynamic stability of the deflagration. In the
realistic case, we see that $\beta$ will split into two parts,
$\beta_\pm$, corresponding to perturbations on each side of the wall.
We remark that, in the case of small supercooling, Eq.
(\ref{fdrsmallsup}) (which depends only on $T_+$) is a good
approximation for the stationary force
$F_{\mathrm{dr}}(\bar{T}_+,\bar{T}_-)$, but not for the general form
$F_{\mathrm{dr}}(T_+,T_-)$, and should not be used to obtain
$b_{\pm}$.

\subsection{Fourier modes of the perturbations}

The fluid equations away from the wall, Eqs. (\ref{HD6}-\ref{HD8}),
can be expressed in  matrix form,
 \beg\label{HD9}
    \left[ \hC_0\pa_0+\hC_z\pa_z+\hC_\pe\pa_\pe  \right]{\vec{U}}=0
 \en
where (from now on we remove the bars on unperturbed variables)
  \beg\label{HD10}
    \hC_0\equiv\left[\begin{array}{ccc}
              1   &  c_s^2w\ga^2 v & 0\\
              v& w \ga^2          & 0\\
          0   &         0         &w\ga^2
             \end{array}\right],\;
   \hC_z\equiv\left[\begin{array}{ccc}
              v   &  c_s^2w\ga^2  & 0\\
              1      &  w\ga^2v   & 0\\
          0      &         0      &w\ga^2v
             \end{array}\right],\;
   \hC_\pe\equiv\left[\begin{array}{ccc}
              0   &  0 & c_s^2w\\
              0   &  0 & 0\\
          1   &  0 & 0
             \end{array}\right],
 \en
and ${\vec{U}}$ is the perturbation vector
 \beg
  {\vec{U}}\equiv\left[\begin{array}{c}
                      \de p\\
              \de v\\
              v^\pe
                     \end{array}\right].
 \en
To solve Eq. (\ref{HD9}) we may use the separation of variables
method which, in this simple case, amounts to searching for solutions
of the form
 \beg\label{Fou1}
    {\vec{U}}(t,z,x^\pe)=\vec{L}e^{\Omega t+qz+ikx^{\pe}},
 \en
where $\Omega, q$ and $ik$ are the eigenvalues of the operators
$\pa_0,\pa_{z}$ and $\pa_{\perp}$, respectively. While $k$ is a real
wavenumber, corresponding to Fourier modes along the wall, $\Omega$
and $q$ are in general complex numbers. The stationary solution will
be unstable whenever $\mathrm{Re} (\Omega)>0$.

Inserting the modes (\ref{Fou1}) into Eqs. (\ref{HD9}) we obtain the system
of homogeneous equations
 \beg\label{Fou3}
  (\hC_0\Omega+\hC_z q+\hC_\pe ik)\vec{L}\equiv C\vec{L}=0.
 \en
The determinant of the matrix $C$ must vanish so that the trivial
solution is not the only one. This gives the dispersion relations
 \beg
 qv+\Omega=0
 \en
or
 \beg \label{Fou4}
    (q v+\Omega)^2-c_s^2(\Omega v+q)^2+c_s^2\ga^{-2}k^2=0,
 \en
and we have three solutions,
 \bega
    {q}_1&=&-{\Omega}/{v} ,\label{Fou10}\\
  q_{2,3}&=&\fr{ (1-c_s^2)v\Omega \pm c_s(1-v^2)\sqrt{\Omega^2+
   ({c_s^2-v^2})\ga^2 k^2}}{c_s^2-v^2}. \label{Fou11}
 \ena
The corresponding eigenvectors are
 \beg\label{Fou7}
      \vec{L}_1=\left[\begin{array}{c}
                      0\\
              1\\
              \fr{iq_1}{k}
                     \end{array} \right],\;
      \vec{L}_{2,3}=\left[\begin{array}{c}
                      -w\ga^2\left(\fr{\Omega+q_{2,3}v}{\Omega v+q_{2,3}} \right)\\
              1\\
              \fr{ik}{\Omega v+q_{2,3}}
                     \end{array} \right]
 \en
The eigenvector $\vec{L}_1$ corresponds to a special solution which
describes isobaric perturbations ($\de p_1=0$) moving with the fluid
(i.e., with $z,t$ dependence of the form $z-v t$).

The general solution is a superposition of these modes. In
particular, for given $k$ and $\Omega$, we must consider a
perturbation vector of the form
 \beg 
    {\vec{U}}(t,z,x^\pe)=\vec{A}(z)e^{(\Omega t+ikx^\pe)},
 \en
with
 \beg\label{Fou2}
    \vec{A}(z)=\sum_{j=1}^3 A_j \vec{L}_je^{q_j z}.
 \en
The function $\vec{A}(z)$ must satisfy the boundary conditions at
$z=\pm \infty$ and the junction conditions at the wall. Accordingly,
the perturbation from the planar shape of the surface will be of the
form
 \beg\label{Fou12}
        \z(t,x^\pe)=D e^{(\Omega t+ikx^\pe )}.
 \en

For  $\mathrm{Re} (\Omega)>0$  the perturbation grows exponentially
with time, whereas for  $\mathrm{Re} (\Omega)<0$ the perturbation
decays exponentially. On the other hand, the condition that the
source is the wall itself, and not something outside it, implies that
the perturbations must decay away from the wall \cite{landau}.
Therefore, instability of the front also requires $\mathrm{Re} (q)<0$
for $z>0$ ($+$ phase) and $\mathrm{Re} (q)>0$ for $z<0$ ($-$ phase).

Since $v$ is negative,  the special solution gives $\mathrm{Re}
(q_1)>0$  for $\mathrm{Re}(\Omega)>0$. Hence, unstable perturbations
will be associated to the presence of this mode in the $-$ phase
(behind the wall). Conversely, for stable perturbations this mode
will be in the $+$ phase.

\begin{figure}[bth] \centering \epsfysize=5.5cm \leavevmode
\epsfbox{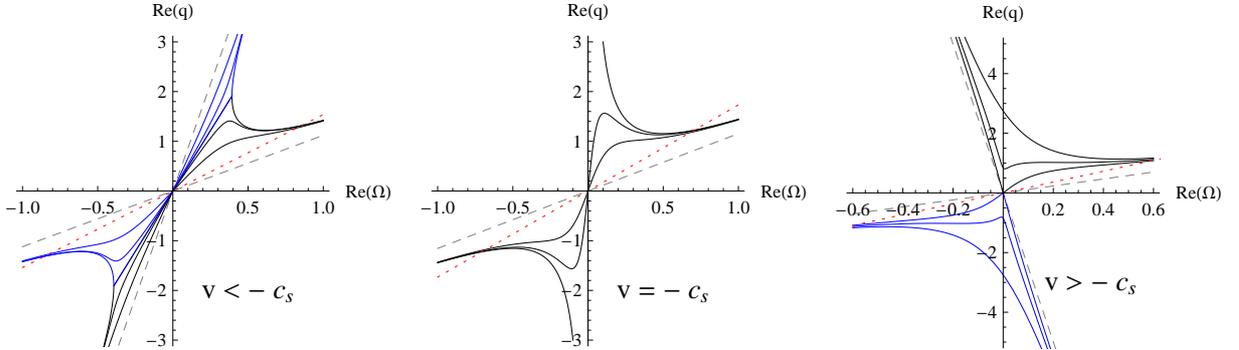} \caption{The real part of the dispersion
relations $q_2(\Omega)$ (in black) and $q_3(\Omega)$ (in blue) for
different values of the imaginary part $\mathrm{Im}(\Omega)$. Gray
dashed lines indicate the asymptotes of these solutions. We also show
the special solution $q_1(\Omega)$ in a dotted red line.}
\label{fighip}
\end{figure}

Regarding the other two solutions, we can write Eq. (\ref{Fou11}) in
the form $(\Omega-a_+q)(\Omega-a_-q)=-K^2$, with
 \beg
 a_\pm=\pm\fr{c_{s}\mp v}{1\mp c_s v}
 \en
and $K^2=c_s^2\ga^{-2}k^2/(1-c_s^2 v^2)$, which shows that, for real
$q$ and $\Omega$, we have a hyperbola with asymptotes
$q=\Omega/a_\pm$. For complex $q$ and $\Omega$, it can be shown that
the points $\mathrm{Re}(q),\mathrm{Re}(\Omega)$ lie in the same
region between these asymptotes. We show some examples in Fig.
\ref{fighip}. We have three different cases, depending on the value
of $v$. For $v$ supersonic (left panel), $\mathrm{Re}(q_2)$ and
$\mathrm{Re}(q_3)$ have the same sign for a given $\Omega$. For $v$
subsonic (right panel), $\mathrm{Re}(q_2)$ and $\mathrm{Re}(q_3)$
have opposite sign. In the case $v=-c_s$ (central panel) there is
only one solution (besides the special one), namely
 \beg\label{Fou6}
    q_2=\fr{c_s k^2}{2\Omega}+\fr{(1+c_s^2)}{2c_s}\Omega.
 \en
In front of the wall we require $\mathrm{Re}(q)<0$, and the possible
modes are those in the lower quadrants of Fig. \ref{fighip}.
Conversely, the possible modes behind the wall are those in the upper
quadrants.

Deflagrations are characterized by  $|v_+|<c_{s+}$, i.e., the
incoming flow is always subsonic. Thus, for the $+$ phase, the right
panel of Fig. \ref{fighip} applies. Furthermore, in this phase we
require $\mathrm{Re}(q)<0$ and the dispersion relation corresponds to
the lower (blue) curves of this panel, i.e., $q=
q_3(v_+,\Omega)\equiv q_{3+}$. In the $+$ phase, the special solution
$q_1$ must be included only for the stable case
$\mathrm{Re}\Omega<0$.

On the other hand, in the $-$ phase we require $\mathrm{Re}(q)>0$,
but the outgoing flow may, in principle, be supersonic as well as
subsonic. As a consequence, the stability analysis is quite different
for weak, strong, or Jouguet deflagrations.

\subsection{Solution of the perturbation equations} \label{solu}

\subsubsection{Supersonic deflagrations}

As we have seen, a supersonic deflagration can be either a strong or
a Jouguet solution. Let us consider first the case of strong
deflagrations. As already discussed by Landau \cite{landau} for a
classical gas, strong deflagrations are absolutely unstable, either
in the case of a combustion front (\S 131) or a condensation
discontinuity (\S 132), due to the fact that such discontinuities are
not \emph{evolutionary} (i.e., there are more free parameters than
conditions). According to numerical calculations \cite{ikkl94},
strong deflagrations are unstable also in the case of a relativistic
phase transition. It is interesting to verify this result
analytically.

Strong deflagrations are characterized by a supersonic outgoing
velocity $v_-$, corresponding to the left panel of Fig. \ref{fighip}.
For $\mathrm{Re}(\Omega) <0$, we see that all the solutions give
$\mathrm{Re}(q)<0$. Therefore, we have no possible modes with
$\mathrm{Re}(\Omega) <0$ in the $-$ phase. On the other hand, for
$\mathrm{Re}(\Omega)>0$ we have $\mathrm{Re}(q_2)>0$,
$\mathrm{Re}(q_3)>0$, and $\mathrm{Re}(q_1)>0$. We thus have three
unstable modes in the $-$ phase. In the $+$ phase, as we have already
discussed,  we have  one unstable mode for $\Omega>0$, namely, $q_3$,
since $\mathrm{Re}(q_3)<0$ (right panel of Fig. \ref{fighip}). Hence,
according to Eq. (\ref{Fou2}), for $z>0$ we must consider a
perturbation of the form
 \beg
 \vec A (z)=A_{+}\vec{L}_{3+} e^{{q}_{3+} z} , \label{modomas}
 \en
whereas for $z<0$ we have
 \beg
\vec{A}(z)=A_-\vec{L}_{3-}e^{{q}_{3-} z}+
B\vec{L}_{2-}e^{{q}_{2-}z}+C\vec{L}_{1-}e^{{q}_{1-}z}.
\label{modosstrong}
 \en
Here, the $\pm$ signs in the $q_i$ and $\vec L_i$  mean that these
quantities [which are given by (\ref{Fou10}-\ref{Fou7})] must be
evaluated at $v_\pm$, respectively. We must impose the junction
conditions (\ref{EM4.1}-\ref{EM4.3}) and the surface equation
(\ref{EM17}) to the fluid variables and the surface deformation
$\zeta$. This gives four equations for the five variables
$A_+,A_-,B,C$, and $D$. Such a system of equations has infinite
solutions. Hence,  the strong deflagration is trivially unstable.
Notice that our treatment applies to any of the strong deflagration
profiles sketched in Figs. \ref{figdefla} and \ref{figstrong}, since
we perturbed the fluid near the wall from constant velocity
solutions.

As we decrease the velocity and reach the Jouguet point
($|v_-|=c_{s-}$), one of the asymptotes becomes vertical and the
hyperbola becomes single-valued (central panel of Fig. \ref{fighip}).
In this case the solution $q_3$ disappears. Thus, we now have only
two unstable modes in the $-$ phase, corresponding to $q_2$ and the
special solution $q_1$. Hence, we must set $A_-=0$ in Eq.
(\ref{modosstrong}). In the $+$ phase the situation is the same as
before (since, for deflagrations, $v_+$ is always subsonic), and the
unstable mode is given by Eq. (\ref{modomas}). As a consequence, the
deflagration becomes evolutionary.

As we have discussed in Sec. \ref{hydrostat}, for the supersonic
Jouguet deflagration the fluid velocity profile develops a tail just
behind the wall (right panel of Fig. \ref{figstrong}), and our
treatment no longer applies. Although it would be interesting to
study the stability of this kind of solution, it is quite a difficult
task since the stationary velocity profile is a function of $z$ and
$t$, namely, $v_\mathrm{rar}(z/t)$. Such a study is out of the scope
of the present paper and we shall attempt it elsewhere.  For the
traditional deflagration profile, the Jouguet point just corresponds
to the case $v_w=c_s$, which is the limit between strong and weak
deflagrations.

\subsubsection{Weak deflagrations}

In the case of weak deflagrations, both velocities $v_-$ and $v_+$
are subsonic. We have $\mathrm{Re}(q_2)>0$ and $\mathrm{Re}(q_3)<0$
(right panel of Fig. \ref{fighip}). Hence, we must consider the mode
with eigenvalue $q_3$ again in the $+$ phase and the mode with
eigenvalue $q_2$ in the $-$ phase. Besides, for
$\mathrm{Re}(\Omega)>0$, we have $\mathrm{Re}(q_1)>0$, and the
special solution must be considered in the $-$ phase. Thus, in the
unstable case, for $z>0$ the function $\vec A(z)$ is again of the
form (\ref{modomas}), $\vec A(z)=A\vec{L}_{3+} e^{{q}_{3+} z}$, while
for $z<0$ we have
$ \vec{A}(z)= B\vec{L}_{2-}e^{{q}_{2-}z}+C\vec{L}_{1-}e^{{q}_{1-}z}$.
The junction conditions (\ref{EM4.1}-\ref{EM4.3}), as well as the
surface equation (\ref{EM17}), require evaluating these functions at
the wall position $z=\z$. However, to first order in the
perturbations, we just evaluate at $z=0$. We thus have, on each side
of the interface (omitting a factor $e^{\Omega t+ik x^\pe}$),
 \beg\label{Weak2}
  \de v_+= A,\;\;
  \de p_+=-\baga_+^2\baw_+\fr{\Omega+{q}_{3+}\bav_+}{\Omega \bav_+
  +{q}_{3+}}A,\;\;
    v^\pe_+=\fr{ik}{\Omega\bav_++{q}_{3+}}A,\;\;
 \en
and
 \beg\label{Weak4}
\de v_-=B+C,\; \de
p_-=-\baga_-^2\baw_-\fr{\Omega+{q}_{2-}\bav_-}{\Omega \bav_-
+{q}_{2-}}B, \; v^\pe_-=\fr{ik}{\Omega\bav_-
      +{q}_{2-}}B+\frac{i{q}_{1-}}{k}C.
 \en
These quantities and those related to the corrugation of the wall,
 \beg
\pa_0\z=\Omega D, \; \pa_{\perp}\z=ikD , \; \pa_0^2\z=\Omega^2 D, \;
\pa_{\perp}^2\z=-k^2D \label{corrug}
 \en
(omitting again a factor $e^{\Omega t+ik x^\pe}$), are related by
Eqs. (\ref{EM4.1}-\ref{EM4.3}) and (\ref{EM17}). We thus have four
equations for the four unknowns $A,B,C$ and $D$, i.e., the weak
deflagration is evolutionary.

It is interesting to consider also the case of a stable perturbation,
$\mathrm{Re}(\Omega)<0$. In this case we have $q_1<0$ and the special
solution must now be included in the $+$ phase instead of the $-$
phase. The form of the perturbations is similar to that of Eqs.
(\ref{Weak2}-\ref{Weak4}), except that the variable $C$ appears in
(\ref{Weak2}) instead of (\ref{Weak4}). As we shall see, the jump of
the special mode from one side of the interface to the other as
$\Omega$ changes sign will cause a discontinuity in the wavenumber
$k$ as a function of $\Omega$.

Inserting Eqs. (\ref{Weak2}-\ref{corrug}) in Eqs.
(\ref{EM4.1}-\ref{EM4.3},\ref{EM17}), we obtain a homogeneous system
of linear equations for the constants $A,B,C$ and $D$. Nontrivial
solutions exist if the determinant of the matrix associated to this
system vanishes. After some manipulations (e.g., multiplying the
first column by the factor $Q_+$ defined below, etc.), this condition
can be written in the form
 \begin{equation}
 \label{matriz}
  \left|
\begin{array}{cccc}
  \frac{1}{R_+} &  \frac{1}{R_-} & \frac{\hat\Omega}{v_-\gamma_-^2} & -(v_+-v_-) \\
  v_-(1+\frac{v_+\W}{R_+}) & -v_+(1-\frac{v_-\W}{R_-}) & v_+(1+v_-^2) & \W (1-v_-v_+)(v_+-v_-) \\
  1+\frac{\W}{v_+R_+} & -1+\frac{\W}{v_-R_-} & 2 & \frac{F_{\mathrm{dr}}}{w_+}\frac{1}{v_+\ga_+^2}(\W^2+1)kd \\
  \fr{\ga_{s+}^2}{2}(\ga_+Q_+-b_+P_+) & \fr{\ga_{s-}^2}{2}(\ga_-Q_--b_-P_-) & -\fr{\ga_-}{2} &
    -\langle\ga\rangle\W+\langle\ga v\rangle(\W^2+1)kd \\
\end{array}%
\right| =0.
  \end{equation}
where $\W\equiv \Omega/k$,  and
\begin{equation}
R_{\pm}=\sqrt{\frac{\gamma_\pm^2}{\gamma_{s\pm}^2}+\frac{\hat{\Omega}^2}{c_{s\pm}^2}},\;\;
P_\pm=v_\pm \mp \frac{\W}{R_\pm}, \;\;
Q_\pm=1\mp \frac{v_\pm}{c^2_{s\pm}}\frac{\W}{R_\pm},
\end{equation}
with $\gamma_{s\pm}\equiv 1/\sqrt{1-v_\pm^2/c_{s\pm}^2}$. Although
the solution is, by symmetry, symmetric in $k$, to obtain these
expressions we have assumed $k>0$\footnote{In particular, we have
inserted a factor of $k$ inside a square root in the expressions for
$q_{2,3}/k$ to obtain the quantities $R_{\pm}$.}. Thus, from now on
we are using the notation $k=|k|$.

A solution of Eq. (\ref{matriz}) is $\W=-\gamma_-v_-$. Indeed, for
this value of $\W$ we have $R_-=\gamma_-$, $P_-=0$, and
$Q_-=\gamma_{s-}^{-2}$. Hence, the second and third columns of the
matrix are proportional, and the determinant vanishes. However, as
explained in Ref.  \cite{hkllm}, this solution is spurious and has no
physical significance, as it leads to vanishing values of the
variables.

Finding analytical solutions for $\Omega(k)$ from Eq. (\ref{matriz}) is a
difficult task. Notice, on the other hand, that the wavenumber $k$ appears
only in the fourth column, in the matrix elements ${34}$ and ${44}$. Hence,
we can readily find an expression for $k$ as a function of $\W$,
 \beg
   kd=\frac{(v_+ -v_-) \left[\det_{14}+\det_{24}(1-v_+v_-)\W\right]-\det_{44}\langle
    \ga\rangle \W}{(1+\W^2)\left[({F_{\mathrm{dr}}}/{w_+})\det_{34}/({v_+\ga_+^2})-\langle\ga
    v\rangle\det_{44}\right]}, \label{kd}
 \en
where $\det_{ij}$ is the determinant of the $3\times 3$ matrix that results
by removing the $i$-th row and the $j$-th column in Eq. (\ref{matriz}).

We remark that, for $\mathrm{Re}(\Omega)<0$, we will have a different
matrix, since the special mode must be considered in the $+$ phase
instead of the $-$ phase. This amounts to changing, in the third
column of the matrix, the indices $\pm\leftrightarrow \mp$ and the
sign of the first three elements.

\section{Stability of weak deflagrations} \label{stab}

We shall now attempt to find all the unstable solutions for a given
wavenumber. Notice that, for real $\Omega$, Eq. (\ref{kd}) facilitates to
study the general properties of the solution. Moreover, one may obtain a
plot of $\Omega$ vs $k$ by just inverting the graph of $kd(\W)$. However, it
is not clear form Eq. (\ref{kd}) whether solutions with
$\mathrm{Im}(\Omega)\neq 0$ are possible as well.

\subsection{Small velocity limit}

In order to understand the general behavior of the function
$\Omega(k)$, it is convenient to consider first non-relativistic
velocities, so that we can write down analytical expressions which
are rather lengthy in the general case. We shall also use, in this
subsection, the approximation of small supercooling, which is
consistent with a small wall velocity and avoids considering a
particular EOS. Thus, from Eq. (\ref{betapm2}) and the general
driving force (\ref{fdr}) we obtain $b_\pm\simeq \langle v\rangle
L/F_{\mathrm{dr}}$. Then we may use, for the stationary driving
force, the approximation (\ref{fdrsmallsup}),
 \beg \label{fdrvcnr}
 F_{\mathrm{dr}}=\frac{\langle v \rangle}{v_+}Lv_c^2,
 \en
where $v_c=\sqrt{1-T_+/T_c}$. We thus obtain
 \beg
 b_\pm ={v_+}/{v_c^2}. \label{bnr}
 \en
Besides, with these approximations Eq. (\ref{EM3.1}) gives $v_--v_+=
v_- L/w_+$ plus higher order in $(T_c-T)/T_c$. Hence, Eq.
(\ref{matriz}) becomes
 \begin{equation}
 \label{matriznr}
  \left|
\begin{array}{cccc}
  \frac{1}{R_+} &  \frac{1}{R_-} & \frac{\hat\Omega}{v_-} & v_- \baL \\
  v_-(1+\frac{v_+\W}{R_+}) & -v_+(1-\frac{v_-\W}{R_-}) & v_+ & -\W v_- \baL \\
  1+\frac{\W}{v_+R_+} & -1+\frac{\W}{v_-R_-} & 2 & (\W^2+1)kd\langle v \rangle\frac{v_c^2}{v_+^2}\baL \\
  \fr{1}{2}(1-\beta_+) & \fr{1}{2}(1-\beta_-) & -\fr{1}{2} &  (\W^2+1)kd\langle v\rangle-\W \\
\end{array}%
\right| =0,
  \end{equation}
where $\baL={L}/{w_+} $, and
 \beg \label{Rnr}
  R_{\pm}=\sqrt{1+\frac{\hat{\Omega}^2}{c_{s\pm}^2}},\;
  \beta_+=\frac{v_+(v_+-\frac{\W}{R_+})}{v_c^2},\;
  \beta_-=\frac{v_+(v_-+\frac{\W}{R_-})}{v_c^2}.
 \en

\subsubsection{Case $|\W|\gg |v_\pm|$}

Let us first look for solutions of large $|\W|$. We are interested in
the instability case $\mathrm{Re} (\W)>0$, for which we have
$R_\pm=\W/c_{s\pm}(1+\mathcal{O}(c_s^2/\W^2))$. Hence, to lowest
order, the parameters appearing in Eq. (\ref{matriznr}) become
 \beg \label{Rapprox}
  \frac{1}{R_\pm}=\frac{c_{s\pm}}{\W}, \;
  \frac{\W}{R_\pm}=c_{s\pm},\;
  \beta_+=\frac{v_+(-c_{s+}+v_+)}{v_c^2},\;
  \beta_-=\frac{v_+(c_{s-}+v_-)}{v_c^2}.
 \en
Notice that, in this limit, we have $\beta_+>0,\beta_-<0$. We can now
easily calculate the $3\times 3$ determinants defined above. The
factors of $\W$ cancel out in $\det_{14}$, whereas
$\det_{24},\det_{34}$, and $\det_{44}$ are of the form
$\det_{i4}=(\W/v_-)\det'_{i4}$, where the $\det'_{i4}$ are
determinants of $\W$-independent $2\times 2$ matrices. Thus, we have
 \beg \label{omegagrande}
\Omega d= \fr{1}{\langle v\rangle }\frac{\det'_{44}+\baL
v_-\det'_{24}}{\det'_{44}- \baL\frac{v_c^2}{v_+^2}\det'_{34}},
 \en

From this expression we see that (in the limit of large $|\Omega/k|$)
$\Omega$ is be real. More importantly, we can see that the rhs of Eq.
(\ref{omegagrande}) is negative, which means that, in fact, we cannot
have $|\W|$ large for $\Omega>0$\footnote{Notice that we are
considering the case $\mathrm{Re} (\W)>0$; for $\mathrm{Re} (\W)<0$,
the matrix in Eq. (\ref{matriznr}), as well as the approximations
(\ref{Rapprox}), would have different forms.}. Indeed, consider for
simplicity $c_{s+}=c_{s-}$. Using the relation $v_-=(1-\baL)v_+$ and
dropping terms $\mathcal{O}(v^2/c_s^2)$, we obtain
 \begin{eqnarray}
 \det{}'_{44}&=&2c_s+\baL(1+c_s^2)|v_-| ,\\
 v_-\det{}'_{24} &=& \frac{c_s}{2}\left[\frac{\baL}{1-\baL}
  +\frac{(2-\baL)|v_-|(c_s-\baL|v_-|)}{v_c^2}\right] ,\\
 - \frac{v_c^2}{v_+^2}\det{}'_{34} &=& \frac{v_c^2}{v_+^2}\frac{2-\baL}{2}|v_-|
  +\frac{\baL(c-\baL|v_-|)}{2(1-\baL)}
  -|v_-|(1+c_s^2),
 \end{eqnarray}
where we have used absolute values to make the signs clearer. We see
immediately that $\det'_{44}$ and $v_-\det'_{24}$ are positive (since
$\baL<1$). On the other hand, in the expression for $-
({v_c^2}/{v_+^2})\det{}'_{34} $, only the last term is negative.
However, in Eq. (\ref{omegagrande}) this term cancels with the last
term of $\det'_{44}$. Hence, since $\langle v\rangle<0$, the rhs of
Eq. (\ref{omegagrande}) is negative. We conclude that there are no
unstable modes with $|\W|\gg v_w$.

\subsubsection{Case $0<|\W|\lesssim|v_\pm|$}

Let us consider now the case of smaller $\W $, up to order $v_\pm$.
To linear order in $v_+$, $v_-$, and $\hat{\Omega}$, we have
$R_\pm=1$, and
 \beg \label{betanr}
  \beta_+=\frac{v_+}{v_c^2}(v_+-\W), \
  \beta_-=\frac{v_+}{v_c^2}(v_-+\W).
 \en
In order to compare with the results of Ref. \cite{hkllm}, we write
down again the determinant in this limit,
 \beg
\left|%
\begin{array}{cccc}
  1 & 1 & \fr{\W}{v_-} & v_- {\baL} \\
  v_- & -v_+ & v_+ & -v_-\W {\baL} \\
  1+\fr{\W}{v_+} & -1+\fr{\W}{v_-} & 2 & kd \frac{v_c^2}{v_+^2}\langle v\rangle {\baL}\\
  \fr{1}{2}(1-\be_+) & \fr{1}{2}(1-\be_-) & -\fr{1}{2} & kd \langle v\rangle -\W\\
\end{array}%
\right|=0. \label{matrnr}
 \en
As expected, the first three rows of the matrix (corresponding to the
junction equations for the fluid perturbations) match\footnote{Taking
into account that $d\,v_c^2/v_+^2=d_c$ and the notations $v_q=-v_+$,
$v_h=-v_-$, $\baL\to\baL/2$.} those of Ref. \cite{hkllm}. All the
differences appear in the forth row (which corresponds to the
equation for the interface). Since we considered independent
perturbations of variables in each phase, this row is more symmetric
in our case. In the case of dependent perturbations, we would replace
$\delta v_-\to\delta v_+$, $\delta p_-\to\delta p_+$ in Eq.
(\ref{EM13}). We would thus obtain zeros in the elements ${42}$ and
${43}$ (corresponding to the perturbations $\de v_-$ and $\de p_-$)
and a factor of $2$ in the element $41$, which would be just given by
$1-\beta$ ($1-\eta$ in the notation of \cite{hkllm}).

As we shall see next, for small velocities, the aforementioned
differences do not introduce a significant qualitative variation with
respect to the results of Ref. \cite{hkllm}. Roughly, the role of
that single $\beta$ will be played by the average of $\be_+$ and
$\be_-$. More important discrepancies will appear for higher
velocities. The results will differ significantly also in the case
$\Omega\leq 0$, even in the non-relativistic case. Indeed, since the
special mode must be considered on either side of the wall according
to the sign of $\Omega$, we cannot use Eq. (\ref{matrnr}) for
$\Omega<0$.  As a consequence, we shall find a discontinuity in the
passage from $\Omega> 0$ to $\Omega<0$.

Let us first consider the case $\mathrm{Re}(\Omega)>0$. Besides the
aforementioned spurious solution $\W=-v_-$, we obtain a solution of
the form
 \beg
  kd=N/D.
 \en
The expressions for $N$ and $D$ are still rather lengthy, and we
shall only write down the case of small $L/w_+$. To first order we
have
 \begin{equation}
 \begin{split}
  kd & \left[ 1+\fr{\baL(1+\fr{\bar{L}}{2})}{2\beta}+\fr{\W}{v_w}  \right] \ = \
   \fr{\baL}{2}\left(1-  \beta\right) \label{kdnrcubic} \\
  & - \left[1+\fr{\baL (1-\baL)\beta}{2}\right]\fr{\W}{v_w}
   -  \left[1+\fr{\baL(1+\beta)}{2}\right]\left[\fr{\W}{v_w}\right]^2
   - \fr{\baL\beta(1+\baL)}{2}\left[\fr{\W}{v_w}\right]^3, 
   \end{split}
 \end{equation}
where we have defined the parameter
 \beg
 \beta=(1-\baL)\; \fr{v_w^2}{v_c^2}. \label{betanrfinal}
 \en
In Eqs. (\ref{betanr}-\ref{betanrfinal}) we have neglected terms of
order $v_w^2$, except in the ratio $v_w^2/v_c^2$, since $v_c$ may be
small. Notice that $v_c^2=1-T_+/T_c$ gives a measure of the amount of
supercooling. Hence, in the non-relativistic approximation, if we are
interested in velocities $v_w\sim v_c$,  the amount of supercooling
must be small enough (of order $v_w^2$), i.e.,
$(T_c-T_+)/T_c=v_c^2\sim v_w^2$.

In order to obtain all the possible values of $\Omega$ (with positive
real part), we should invert the relation (\ref{kdnrcubic}), which
amounts to finding the three roots of a cubic polynomial. Notice that
the coefficient of the cubic term is in general suppressed with
respect to the the linear and quadratic ones. If we neglect this
term, we obtain a quadratic equation, with only two roots. The third
root of the cubic equation must be large enough to make the cubic
term comparable to the quadratic one, i.e., $\W/v_w\sim \baL^{-1}$.
This root is in general well beyond the range of validity of the
approximation $\W\lesssim v_w$ and must be discarded (in any case, it
can  be seen that this solution  has a negative real part). Let us
just assume, for simplicity, that $\baL$ is small enough that we can
neglect the cubic term. Then, we have a quadratic equation with real
coefficients, and it is trivial to see the structure of the
solutions. We have either two complex conjugate roots or two real
roots. Since the coefficients of the quadratic and linear terms have
the same sign, in the case of complex roots the real part is
negative. In the case of real roots, one of them is negative. The
other has a smaller absolute value and may be positive or negative,
depending on the sign of the $\W$-independent term. Thus, we have, at
most, only one solution with $\mathrm{Re}(\Omega)>0$. This solution
has $\mathrm{Im}(\Omega)=0$.

To study this solution, it is convenient to analyze the behavior of
$k$ as a function of $\Omega$. We can thus go back to Eq.
(\ref{kdnrcubic}) and consider $\Omega$ real.  For $\W>0$ the lhs of
Eq (\ref{kdnrcubic}) is positive, while most of the terms in the rhs
are negative. Indeed, the coefficients of the $\W$-dependent terms
are negative. The $\W$-independent term is negative for $\beta>1$. In
such a case, there is no solution with $\W> 0$ (i.e., we would have
unphysical values $k<0$), and the deflagration is stable under
perturbations of any wavelength. For $\beta<1$, the $\W$-independent
term in the rhs is positive and we have unstable solutions for $\W$
below a certain value $\W_{0}$. At $\W=\W_0$ we have $k=0$ and, as
$\W$ decreases from $\W_{0}$, the value of $k$ increases. For
vanishing $\W$ we obtain the maximum wavenumber $k_c$ for which the
perturbation is exponentially unstable.

\subsubsection{Analytic approximations} \label{analyt}

In the limit of very small $\baL$, the critical value $\beta =1$ is
attained for $v_w=v_c$. This is in agreement with Ref. \cite{hkllm}
(in this limit, and for $\W\to 0$, we have $\beta_+=\beta_-=\beta$).
If $\baL$ is not negligible, the critical velocity is somewhat
higher,
 \beg
 v_{\mathrm{crit}}=v_c/\sqrt{1-\baL}\approx v_c(1+\baL/2).
 \label{vcL}
 \en
For $v_w$ below $v_{\mathrm{crit}}$, we have $\Omega>0$ in the range
$0<k<k_c$. The value of $k_c$ is obtained by setting $\W=0$ in Eq.
(\ref{kdnrcubic}),
 \beg \label{kcdlinear}
 k_cd=\frac{\baL}{2}\frac{1-\beta}{1+(1+\fr{\baL}{2})\fr{\baL}{2\beta}}.
 \en
This equation shows explicitly the fact that all perturbations are
stable for $\beta>1$.

\begin{figure}[bth]
\centering
\epsfysize=5.5cm \leavevmode \epsfbox{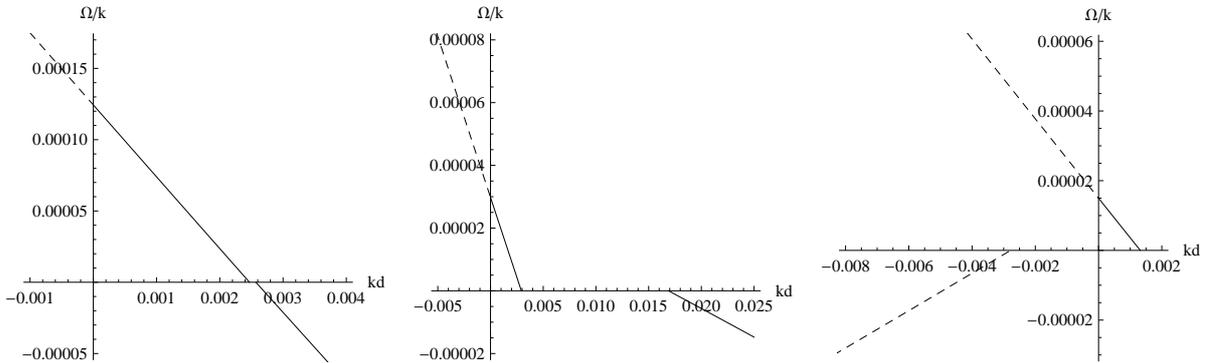}
\caption{$\W$ vs $kd$ in the  small $v_w$, small $L$ and small $\W/v_w$ approximation,
for $\baL=0.01$, $T_+/T_c=0.995$ ($v_c\simeq 0.07$), and $v_w=0.05$ (left), $0.007$ (center)
and $0.003$ (right).}
\label{figlinear}
\end{figure}

As already mentioned, to obtain the value of $\Omega$ for a given
$k$, we should in principle invert the cubic equation
(\ref{kdnrcubic}). In Ref. \cite{hkllm} the corresponding equation is
quadratic  because the dependence of the parameter $\beta_+$ on
$\Omega$ is neglected. Furthermore, it is argued that the smallness
of the velocity (and the fact that $\W\sim v_w$) insures that the
term quadratic in $\Omega$ is small and can also be neglected,
obtaining a linear equation. Notice, however, that $\W\sim v_w$ is
not a good argument to neglect any of the terms in the equation.
Nevertheless, since the independent term in the rhs of Eq.
(\ref{kdnrcubic}) is always smaller than $\baL/2$, we have
$\W_{0}/v_w< \baL/2$. Since we always have $\baL<1$ (and often $\ll
1$), this is an important constraint. As a consequence, for the
unstable range $0<\Omega<\Omega_{0}$, the value of $\W/v_w$ will be,
in most cases, small enough to safely neglect the quadratic and cubic
terms in Eq. (\ref{kdnrcubic}). Keeping only the linear terms we
obtain
 \beg \label{omegalinear}
 \fr{\W}{v_w }=\frac{\baL}{2}\frac{(1-\beta)(1-k/k_c)}{1+(1-\baL)
 \fr{\baL}{2}\beta+kd},
 \en
which agrees with the result of Ref. \cite{hkllm} for $\baL\ll 1$.
Thus, we have
 \beg \label{omegamaxlinear}
 \fr{\W_{0}}{v_w }=\frac{\baL}{2}\frac{1-\beta}{1+(1-\baL)\fr{\baL}{2}\beta}
 \en
We see that both $\W/v_w$ and $kd$ are at most of order $\baL$. We
plot the function $\W$ in Fig. \ref{figlinear}.

Notice that Eq. (\ref{omegalinear}) is valid only for $\W>0$.
Although we are not interested in general in the case
$\mathrm{Re}(\Omega)<0$, which is exponentially stable, it is
important to consider the limit $\Omega\to 0^-$. As we have seen, for
$\mathrm{Re}(\Omega)<0$ the special  mode must be considered in front
of the wall. This results in a change in the third column of the
matrix in Eq. (\ref{matrnr}), which becomes
$\mathrm{col}(-\W/v_+,-v_-,-2,-1/2)$. This will give a jump in $k$ as
a function of $\Omega$ (see Fig. \ref{figlinear}). Proceeding as
before we obtain, for small (and negative) $\W/v_w$,
 \beg \label{omegalinearneg}
 \fr{\W}{v_w}=
\frac{\baL}{2}\frac{(1-\beta+\frac{\baL}{2})(1-k/k_c')}{1-
(1+\baL)\fr{\baL^2}{2}\beta-(1+2\baL)kd} \;\;\; (\Omega<0),
 \en
where the minimum wavenumber for which the perturbation is
exponentially stable is given by
 \beg \label{kcdlinearneg}
 k_c'd=\frac{\baL}{2}\frac{1-\beta+\baL}{1-(1+\fr{3\baL}{2})\fr{\baL}{2\beta}}.
 \en

Comparing Eq. (\ref{omegalinearneg}) with Eq. (\ref{omegalinear}), we
observe some differences of order $\baL \sim\Delta v/v$ in the
expressions for $\Omega>0$ and $\Omega<0$ (due to the changes $v_\pm
\leftrightarrow v_\mp$ in the third column of the matrix).  On the
other hand, from Eq. (\ref{kcdlinearneg}) we see that there is also
an important sign difference in the denominator of $k_c'$ with
respect to that of $k_c$.  For $\beta\sim 1$ (i.e., $v_w\sim v_c$),
$k_c'$ will be slightly higher than $k_c$. As a consequence, there
will be a small gap in the plot of $\Omega$ vs $k$, as can be seen in
the left panel of Fig. \ref{figlinear}. This gap grows significantly
as $v_w$ decreases from $v_c$ (center panel of Fig. \ref{figlinear}),
since $k_c'$ has a pole at $\beta\approx \sqrt{\baL/2}$. Hence, for a
wall velocity $v_c'\approx (\baL/2)v_c$ the interval  $k_c<k<k_c'$
becomes infinite. Below this velocity, $k_c'$ takes negative values
(right panel of Fig. \ref{figlinear}) and the solution with
$\Omega<0$ becomes unphysical. There are, in general, more solutions
with $\mathrm{Re}(\Omega) <0$, for higher values of $|\Omega|$. We
are not interested in them, though, since they correspond to stable
perturbations. In this case, we have linear stability for $k>k_c$.

\begin{figure}[bth]
\centering
\epsfysize=4.5cm \leavevmode \epsfbox{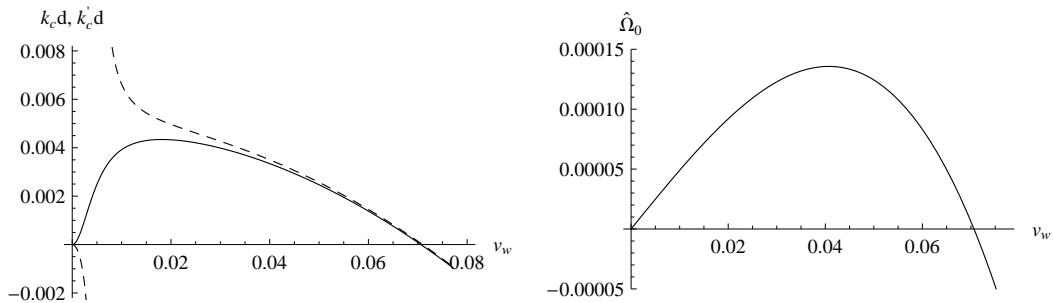}
\caption{The values of $k_c d$ (solid line), $k_c'd$
(dashed line), and $\W_{0}$ as functions of $v_w$,
for the same set of parameters of Fig. \ref{figlinear}.}
\label{figkcommaxnr}
\end{figure}

In the region of $\Omega>0$ our results agree with those of Ref.
\cite{hkllm}. In this region, $\W$ decreases from $\W_{0}$ to $0$ as
$k$ increases from $0$ to $k_c$. Thus, the instability range of $k$
is limited by $k_c\lesssim \baL/2$, whereas $\W$ is bounded by
$\W_{0}\lesssim v_w\baL/2$. These values are also proportional to
$1-v_w^2/v_{\mathrm{crit}}^2$. Therefore, for higher velocities they
are even smaller, and stability is recovered at
$v_w=v_{\mathrm{crit}}$. According to Eqs. (\ref{kcdlinear}) and
(\ref{omegamaxlinear}), for small velocity we have $k_cd\sim v_w^2$
and $\W\sim v_w$. Hence, stability is also recovered as $v_w$
vanishes. This is shown in Fig. \ref{figkcommaxnr} (solid lines). For
a given velocity, the exponentially unstable wavenumbers are those
below the curve of $k_c$, whereas the possible values of $\W$  lie
below the curve of $\W_{0}$. Beyond the critical velocity both $k_c$
and $\W_{0}$ become negative.

For $k\geq k_c$, on the other hand, we have significant differences
with the results of Ref. \cite{hkllm}. The gap between $k_c$ and
$k_c'$  was not observed in that analysis. The reason may be the
following. As we have seen, for $\Omega<0$ the only changes in the
matrix in Eq. (\ref{matrnr}) are in the third column. For $\W=0$, the
discontinuity in this column is given by the change
$\mathrm{col}(0,v_+,2,-1/2)\to \mathrm{col}(0,-v_-,-2,-1/2)$. Since
changing the sign of a column does not alter Eq. (\ref{matrnr}), the
above change is equivalent to
$\mathrm{col}(0,v_+,2,-1/2)\to\mathrm{col}(0,v_-,2,1/2)$. In the case
of small $\baL$, the change $v_+\to v_-$ is not relevant. Therefore,
the discontinuity is dominated by the change of sign $-1/2\to 1/2$ in
the element 43. However, due to the simplified treatment of Ref.
\cite{hkllm}, in that work this element is $0$ instead of $-1/2$.

In principle, at $k=k_c$, $\Omega$ jumps to a value $\Omega<0$ (not
shown in Fig. \ref{figlinear}, as it lies beyond the linear
approximation). It is not clear, though, from the linear stability
analysis, whether there are unstable perturbations or not in the
range $k_c<k<k_c'$. For $k<k_c$ the perturbations grow exponentially,
whereas for $k>k_c'$ they decay exponentially. In the range between
$k_c$ and $k_c'$, one may expect marginal stability with $\Omega=0$.
Notice  that, as $\Omega\to 0$, the special solution, which is
proportional to $\exp(\Omega z/|v|)$, becomes non-normalizable in
either side of the wall, and cannot be included at all in Eqs.
(\ref{Weak2}-\ref{Weak4}). As a consequence, we will have four
equations for the three unknowns $A,B,D$, and the only solution will
be the trivial one, $A=B=D=0$. Thus, the approximation of keeping to
linear order in perturbations breaks down. It is out of the scope of
the present paper to go beyond the linear stability analysis. In any
case, in the range of wavenumbers between $k_c$ and $k_c'$, the
perturbations will not grow exponentially. For our purposes, it will
be enough to assume that, in this range, the possible instabilities
would grow more slowly than for $k<k_c$. From now on, we shall
concentrate on the case of $\Omega>0$.

\subsubsection{Reheating effects}

The stability of the perturbations depends on the wall velocity and
on the temperature $T_+$. Notice, however,  that the boundary
conditions for the fluid fix the temperature beyond the shock front
(the nucleation temperature $T_N$), while $T_+$ depends on the amount
of reheating (see Fig. \ref{figdefla}). For a given nucleation
temperature $T_N$, the temperature $T_+$ will depend on the wall
velocity. As a consequence, $T_+$, as well as $v_w$, depend on $T_N$
and the friction. For the stability analysis it is useful to
eliminate the friction and use the wall velocity as a free parameter.
However, it is not reasonable to regard $T_+$ and $v_w$ as free
independent parameters.  In particular, some combinations of $T_+$
and $v_w$ will be unphysical.

For small supercooling (i.e., $T_N$ close to $T_c$), the reheating is
given by \cite{ms09}
 \beg
  \frac{T_+}{T_c}=\frac{T_N}{T_c}+\frac{L}{\sqrt{3}w_+}v_w.
  \label{rehnr}
 \en
In the case $L/w_+\ll 1$ this gives $T_+\approx T_N$. However, a
small temperature variation may cause important effects on the wall
dynamics. Using Eq. (\ref{rehnr}) we can write the wall velocity
(\ref{vwsmallsup}) in terms of $T_N$,
 \beg
   v_w=\fr{w_+}{w_-} \frac{L(1-T_N/T_c)}{\eta_{\mathrm{eff}}}, \label{vwaproxtn}
 \en
where we have defined an effective friction coefficient which takes into
account the reheating in front of the wall \cite{ms09}. We have
 \beg\label{etaeff}
  \frac{\eta_{\mathrm{eff}}}{L}= \frac{\eta}{L}+\frac{L}{\sqrt{3}w_-}.
 \en
Notice that the effects of reheating will depend on the two ratios $\eta/L$
and $L/w_-\sim\baL$, whose values are quite unrelated. Thus, the effective
friction coefficient $\eta_{\mathrm{eff}}$ may be considerably larger than
$\eta$ (even for small $\baL$). In particular, for vanishing $\eta$ we will
still have a finite effective friction. This hydrodynamic obstruction to the
wall motion was discussed more recently in Ref. \cite{kn11}. Notice that,
for $T_+$ fixed, the velocity would only be bounded by relativity for
$\eta\to 0$ [see Eq.(\ref{vwsmallsup})]. In contrast, for $T_N$ fixed,
according to Eqs. (\ref{vwaproxtn}-\ref{etaeff}), the velocity may be
bounded by a relatively low value. As a consequence, some velocities will be
unreachable, as they would require a negative $\eta$.

Similarly, the velocity $v_c$ can be written in terms of $T_N$,
 \beg
  v_c^2=\frac{T_c-T_N}{T_c}\frac{\eta}{\eta_{\mathrm{eff}}}. \label{vcn}
 \en
As we have mentioned, in Ref. \cite{hkllm}, the approximation $T_+=
T_N$ was used. Fixing $T_N$ and $v_w$ instead of $T_+$ and $v_w$, the
stability parameter $\beta\approx v_w^2/v_c^2$ is enhanced, with
respect to that approximation, by a factor of
$\eta_{\mathrm{eff}}/\eta$. The ratio
 \beg
  \frac{\eta_{\mathrm{eff}}}{\eta}=1+\fr{1}{\sqrt{3}}\frac{L}{w_-}\fr{L}{\eta}
 \en
can make a difference if $L/\eta$ is large. Thus, for small
velocities (i.e., large $\eta$) the approximation $T_+=T_N$ will not
be too bad. Notice, on the other hand, that this is a stabilizing
effect. The factor $\eta/\eta_{\mathrm{eff}}$ in Eq. (\ref{vcn})
opposes to the increase with $\baL$ in Eq. (\ref{vcL}). For high
velocities the approximation $T_+=T_N$ will fail as well as the
approximation $T_-=T_+$.

\subsection{Arbitrary velocities} \label{arbv}

The previous analytic treatment may be extended beyond the limits of the
above approximations (e.g., by considering higher orders in $v_w$, $\W$, or
$\baL$). However, the equations are rather lengthy to write down here. We
have also explored the solutions of Eq. (\ref{matriz}) in all the regions of
parameter space. We found that most of the qualitative features of $\W(k)$
hold in the whole range $0<v_w<c_{s}$.  Thus, the only solution with
$\mathrm{Re}(\W)>0$ is real and is always bounded by the value $\W_{0}$
corresponding to $k=0$. For $k>0$ the value of $\W$ decreases. In general,
$\W$ vanishes at a finite value $k_c$, and we have $\W_{0}\lesssim v_w\baL$
and $k_cd\sim\baL$ (there are some exceptions, though; see the next
section). Beyond $k_c$, there may be a range $k_c<k<k_c'$ of marginal
stability. For $k>k_c'$ the perturbations are exponentially stable. The
general behavior of $k_c'$ is qualitatively similar to that observed
analytically. We shall be interested mostly in the case of exponentially
unstable perturbations $k<k_c$.\footnote{A characteristic feature is, thus,
that the instability range appears continuously below a critical velocity.
Furthermore, $\W$ grows continuously below $k=k_c$, and is in general small.
In contrast, in the case of detonations, instabilities generally arise,
below a critical velocity, with large values of $\W$ for all wavenumbers
\cite{stabdeto}.}

Setting $\W=0$ in Eq. (\ref{kd}) we obtain the critical wavenumber,
 \beg \label{kcd}
   k_c d=\frac{\Delta v \, \det_{14}^0}{({F_{\mathrm{dr}}}/{w_+})
   \det_{34}^0/({v_+\ga_+^2})-\langle\ga
    v\rangle\det_{44}^0} .
 \en
Setting $k=0$ we obtain, to linear order in $\W$,
 \beg \label{ommax}
   \W_{0} \simeq\frac{\Delta v \, \det_{14}^0}{\langle\ga
   \rangle\det_{44}^0-\Delta
   v\left[\det_{24}^0(1-v_+v_-)+\det_{14}^1\right]}.
 \en
We have used the notations
$\det_{ij}=\det_{ij}^0+\det_{ij}^1\W+\mathcal{O}(\W^2)$. We write
down, as an example, the determinant  $\det_{14}^0$ (i.e.,
$\det_{14}$ evaluated at $\W=0$),
\begin{equation}
 \det{}_{14}^0=\Delta v \left[\gamma_{s-}^2(\gamma_--v_-b_-)-
 \fr{\gamma_-}{2}\right]+\fr{-v_+}{\gamma_-^2}\left\langle\gamma_{s}^2(\gamma-vb)\right\rangle ,
 \label{det14}
\end{equation}
where $b_\pm$ are the stability parameters defined in Eq.
(\ref{betapm2}). This determinant dominates the behaviors of $k_c$
and $\W_{0}$. In particular, it can be seen that the denominators in
Eqs. (\ref{kcd}) and (\ref{ommax}) are always positive. Thus, the
signs of $k_c$ and $\W_{0}$ depend essentially on the factors
$\gamma_{\pm}-v_{\pm}b_{\pm}$ in Eq. (\ref{det14}) (notice that
$\Delta v$ and $-v_+$ are positive). Hence, the stability is
dominated by the quantities $1- \beta_{\pm}$, with
 \beg \label{betapmfinal}
 \beta_\pm =v_\pm b_\pm/\gamma_\pm .
 \en
These definitions of $\beta_\pm$ are essentially the same as in the previous
subsection, but evaluated at $\W=0$ [cf. Eqs. (\ref{Rnr}),(\ref{betanr})].
The denominators in Eqs. (\ref{kcd}) and (\ref{ommax}) become important for
$v_w$ close to $c_{s-}$. Indeed, in the limit $v_-\to c_{s-}$ we have
$\gamma_{s-}\to \infty$, and $\det_{14}^0$ diverges. This divergence is
canceled by factors of $\gamma_{s-}$ appearing in the denominators. In any
case, $\W_{0}$ and $k_c$ are still dominated by the quantities
$1-\beta_\pm$, which appear in all the determinants.

\subsubsection{High velocities and the Jouguet point} \label{highv}

In the non-relativistic, small $\baL$ case, we have $v_\pm\simeq
v_w$, $b_\pm \simeq v_w/v_c^2$ [cf. Eq. (\ref{bnr})]. As a
consequence, $k_c$ and $\W_{0}$ are proportional to $1-\beta\simeq
1-v_w^2/v_c^2$. This simple expression is a consequence of the fact
that the driving force is proportional to $v_c^2=1-T_+/T_c$ [cf. Eq.
(\ref{fdrvcnr})]. In the general case, it is always possible to
define a velocity $v_c$ which is proportional to the driving force
(hence, $v_c$, as well as $F_{\mathrm{dr}}$, will vanish for
$T_-=T_+=T_c$). Thus, according to Eq. (\ref{betapm2}), the
quantities $b_{\pm}$ will be of the form $\langle\gamma v\rangle
/v_c^2$, and we have
 \beg
\beta_\pm\approx\fr{\gamma_\pm^{-1}\langle\gamma v\rangle v_\pm
}{v_c^2} .
 \en
Notice that we have $\beta_\pm>0$. In the case $v_+\simeq v_-$, we
obtain $\beta_\pm\approx v_w^2/v_c^2$, and the behavior is similar to
the non-relativistic case (namely, $\W_{0}$ will become negative for
a velocity $v_{\mathrm{crit}}\approx v_c$). However, for large $v_w$
we may have a relatively large difference between $v_+$ and $v_-$.

For a deflagration we have $|v_+|<|v_-|=v_w$ and, consequently,
$\beta_+<\beta_-$. Therefore, the factor $1-\beta_-$ vanishes for a
certain velocity $v_w\gtrsim v_c$, but $1-\beta_+$ remains positive
until $v_w$ is increased further. As a consequence, the critical
velocity $v_{\mathrm{crit}}$ (at which $\W_0$ and $k_c$ vanish) will
be higher than $v_c$. For large $\Delta v$, we may have
$v_{\mathrm{crit}}$ close to $c_{s-}$ for relatively low values of
$v_c$. Moreover, $|v_+|$ is bounded by a subsonic value
$v_J^{\mathrm{def}}$. If $v_c$ is higher than this value, then
$1-\beta_+$ may be positive in the whole range $0<v_w<c_{s-}$. Then,
it may happen that $\W_{0}$ never becomes negative, i.e., that there
is no critical velocity at all. In such a case (which will depend on
the amount of supercooling), the deflagration will be unstable for
any subsonic velocity. Moreover, as we shall see in the next section,
the values of $\W_0$ and $k_c$ may become large as $v_w$ approaches
the speed of sound.

This result is in clear contradiction with Ref. \cite{hkllm}, where
it is claimed that it is possible to show that, in the limit $v_w\to
c_s$, the equation for $\Omega$ has no positive roots, for any value
of $k$. This discrepancy is, probably, due to the approximations
$v_+=v_-=v_w$, $T_-=T_+$ used in \cite{hkllm} for the interface
equation. Physically, the stability found for the weak deflagrations
in this limit is explained in Ref. \cite{hkllm} by the fact that the
result matches with the stability of detonations. However, weak
deflagrations never match detonations, as the latter have higher,
supersonic velocities $v_w\geq v_J^{\det}$. Between the speed of
sound and the  Jouguet detonation velocity $v_J^{\det}$, we may have,
in principle, either strong deflagrations or Jouguet deflagrations.
As we discussed in Sec. \ref{stationary}, both match the weak
deflagration at $v_w=c_{s-}$ (i.e., the hydrodynamic solution
bifurcates at the Jouguet point). As we have seen, the strong
deflagration is unstable, whereas the supersonic Jouguet deflagration
is presumably stable in general.

Regardless of the behavior for $v_w\to c_{s-}$, it is easy to show
that \emph{there cannot be a solution with $\Omega<0$ for
$v_w=c_{s-}$}. Indeed, since $v_-=-c_{s-}$, Fig. \ref{fighip}
(central panel) shows that, for $\Omega<0$, all the modes have $q<0$.
Thus, we have no mode behind the wall. In front of the wall, we have
$v_+>-c_{s+}$ (right panel), and we see that there are two modes
($q_1$ and $q_2$) with $q>0$. Applying the linear perturbation
analysis, we will have only three unknowns (namely the amplitudes of
these two modes and that of the surface deformation) for our four
equations (\ref{EM4.1}-\ref{EM4.3},\ref{EM17}). This means that the
analysis of linear perturbations breaks down\footnote{For supersonic
Jouguet deflagrations this argument does not apply, since we are
considering perturbations from a constant velocity, while this
solution has a rarefaction wave immediately after the wall.} for
$\Omega<0$.

For $\Omega>0$, in contrast, the calculation is similar to the
subsonic case (cf. the center and right panels of Fig. \ref{fighip}),
only we must use Eq. (\ref{Fou6}) for $q_2$ instead of Eq.
(\ref{Fou11}). We have checked that the result of such calculation
matches the result of the subsonic calculation  in the limit $v_w\to
c_s$.

\section{Numerical results} \label{numres}

\subsection{The Bag equation of state}

To proceed to the calculation of $\Omega(k)$, we need to consider a
concrete equation of state. The simplest phenomenological model for a
phase transition is the bag EOS, which consists of radiation and
vacuum energy densities (see, e.g., \cite{s82}). The pressure in each
phase can be written in the form
 \beg
 p_+(T)=\fr{a}{3}T^4-\fr{L}{4},\;\;   p_-(T) =\left(\fr{a}{3}-\fr{L}{4T_c^4}\right)T^4.
 \en
The entropy and enthalpy densities can be obtained from
$s=dp/dT,w=Ts$. This model depends on three parameters, namely, the
critical temperature $T_c$, the latent heat $L$, and the coefficient
$a$. The latter is related to the number of effective massless
degrees of freedom in the $+$ phase. The simplicity of the model
often allows to obtain analytic results. The speed of sound is the
same in both phases $c_{s\pm}=1/\sqrt{3}\equiv c_s$.

The solution for the wall velocity can be obtained from Eq.
(\ref{solfric}), $\eta\lan\gamma v\ran=-F_{\mathrm{dr}}$, using the
matching conditions (\ref{EM3.1}-\ref{EM3.2}) and the boundary
conditions. Since the pressure in both phases is a function of $T^2$,
it is convenient to use Eq. (\ref{fdr2}) for the driving force. We
obtain
 \beg
 F_{\mathrm{dr}}=\fr{L}{4}\left(1-\fr{T_-^2T_+^2}{T_c^4}\right).
 \label{fdrbag}
 \en
The matching conditions give the relations
 \beg
 \fr{T_-^2}{T_+^2}=\sqrt{\fr{v_+ \gamma_+^2}{v_- \gamma_-^2 (1 -
 \baL)}},
 \en
\begin{equation} \label{steinhardt}
v_{+}=\frac{1}{1+\alpha_+ }\left[
\frac{1}{6v_{-}}+\frac{v_{-}}{2}\pm \sqrt{\left( \frac{1}{6v_{-}}%
+\frac{v_{-}}{2}\right) ^{2}+\alpha_+ ^{2}+\frac{2}{3}\alpha_+ -\frac{1}{3}}\right],
\end{equation}%
where
 \beg
 \baL\equiv \fr{L}{4aT_c^4/3}=\fr{L}{w_+(T_c)},
 \en
and
 \beg
 \alpha_+\equiv \fr{L}{4aT_+^4}=\fr{\baL}{3}\fr{T_c^4}{T_+^4}.
 \en
The $+$ sign in Eq. (\ref{steinhardt}) corresponds to detonations,
and the $-$ sign to deflagrations. For weak deflagrations, we have
$v_-=-v_w$, and the reheating temperature $T_+$ is related to the
nucleation temperature $T_N$ by
\begin{equation}
\frac{\sqrt{3}\left( T _{+}^4-T_N^4\right) }{\sqrt{\left( 3T_+^4+T_N^4\right)
\left( 3T_N^4 + T_+^4\right) }}
=\frac{v_{+}-v_{-}}{1-v_{+}v_{-}}
.  \label{tmatn}
\end{equation}

We define the velocity $v_c$ by
 \beg
v_c^2 \equiv \fr{1}{4}\left(1-\fr{T_-^2T_+^2}{T_c^4}\right)
 \label{vcbag}
 \en
so that we have $ F_{\mathrm{dr}}=L v_c^2 $. The velocity $v_c$ is
symmetric in $T_+$ and $T_-$. For small supercooling we have
$v_c^2\simeq \fr{1}{2}(1-T_+T_-/T_c^2)\simeq 1-\sqrt{T_+T_-}/T_c$.
For small latent heat we have $T_-\simeq T_+$ and we recover the
definition $v_c^2= 1-T_+/T_c$. From Eqs. (\ref{betapm2}) and
(\ref{fdrbag}) we see that the coefficients $b_\pm$ are equal,
 \beg \label{bbag}
 b_\pm=\fr{\langle\gamma v\rangle}{v_c^2}\fr{T_+^2T_-^2}{T_c^4},
 \en
and the quantities $\beta_\pm$ defined in (\ref{betapmfinal}) are
given by
 \beg \label{betabag}
 \beta_\pm=\fr{\gamma_\pm^{-1}\langle\gamma v\rangle
 v_\pm}{v_c^2}\fr{T_+^2T_-^2}{T_c^4}.
 \en
For small supercooling and small latent heat, we have
$\beta_\pm\simeq  v_w^2/v_c^2$, and we have a critical velocity
$v_{\mathrm{crit}}=v_c$. In the general case, the temperature ratios
in Eq. (\ref{betabag}) enhance the value of $v_{\mathrm{crit}}$ with
respect to $v_c$. Furthermore, the fact that $|v_+|<|v_-|$ implies
that $\beta_+<\beta_-$, as already discussed.

\subsection{Stability of deflagrations}

\begin{figure}[bth]
\centering
\epsfysize=5cm \leavevmode \epsfbox{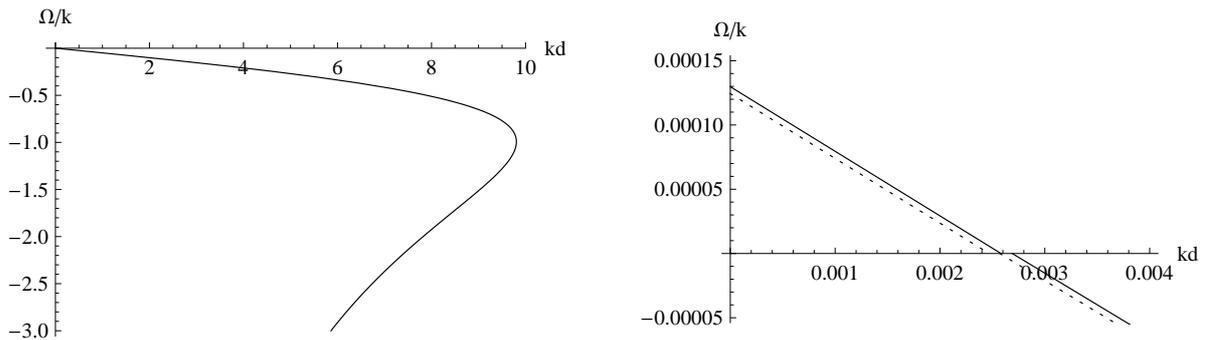}
\caption{$\W$ vs $kd$ for $\baL=0.01$, $T_+/T_c=0.995$, and $v_w=0.05$.
The right panel shows also the linear approximations (dotted line).}
\label{figapps}
\end{figure}

In Fig. \ref{figapps} we plot the set of real solutions for $\W$ as a
function of $kd$ (the right panel zooms at small $kd$). For
$\mathrm{Re}(\W)>0$, we have found no other solutions (neither real
nor complex). We considered values of the parameters similar to those
considered in Ref. \cite{hkllm}, namely, a very small value of the
latent heat ($\baL=0.01$) and a value of $T_+$ very close to $T_c$,
which gives a small\footnote{See the discussion below Eq.
(\ref{betanrfinal}).} critical velocity, $v_{\mathrm{crit}}\simeq
v_c\simeq 0.07$. We have chosen a wall velocity below the critical
one, $v_w=0.05$, so that there is a range of unstable wavenumbers. In
the right panel we have also plotted the approximations
(\ref{omegalinear}) and (\ref{omegalinearneg}). Notice that these do
not match the exact solution even for vanishing $\W$. This is because
the approximations are linear not only in $\W$, but also in the
parameters $v_w,v_c$ and $\baL$.

\begin{figure}[bth]
\centering
\epsfysize=5cm \leavevmode \epsfbox{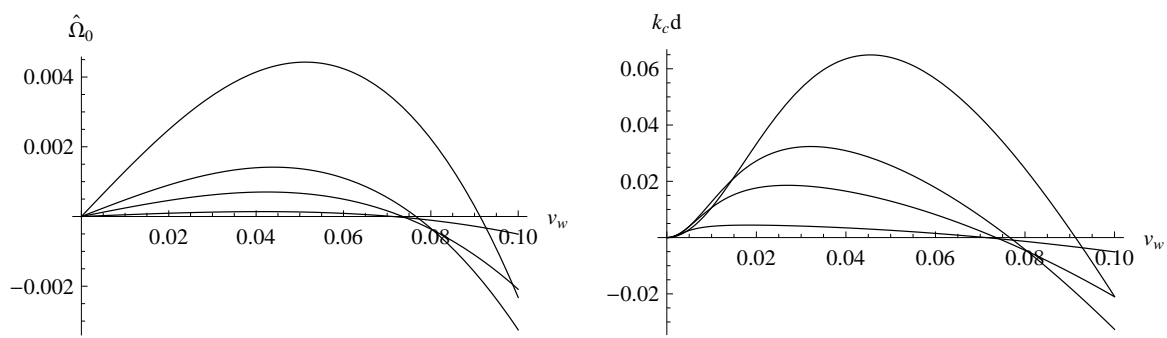}
\caption{The maximum value of $\Omega/k$ (corresponding to $k=0$)
and the maximum
unstable wavenumber $k_c$ (corresponding to $\W\to 0^+$),
as functions of the wall velocity, for $T_+/T_c=0.995$ and, from
bottom to top, $\baL=0.01$, $0.05$, $0.1$ and $0.3$.} \label{figkcommax}
\end{figure}

Changing the values of the parameters, the behavior is qualitatively
similar (as we already discussed analytically). Essentially, the
effect will be a variation of the points where the curves cut the
axes, i.e., of the parameters $\W_0$, $k_c$, and $k_c'$ (see Fig.
\ref{figlinear}). In Fig. \ref{figkcommax}, the two parameters which
characterize the instability (namely, $k_c$ and $\W_{0}$) are plotted
as functions of the wall velocity. The lower curve corresponds to the
parameters of Fig. \ref{figapps} and is well approximated by the
non-relativistic approximation shown in Fig. \ref{figkcommaxnr}. The
other curves in Fig. \ref{figapps} correspond to higher values of the
latent heat. As already seen with analytic approximations, the
critical velocity increases with the latent heat. Notice, however,
that this effect is quite small. Although $k_c$ and $\Omega_{0}$ are
proportional to $\baL$, the critical velocity hardly varies with
$\baL$.

\begin{figure}[bth]
\centering
\epsfysize=5cm \leavevmode \epsfbox{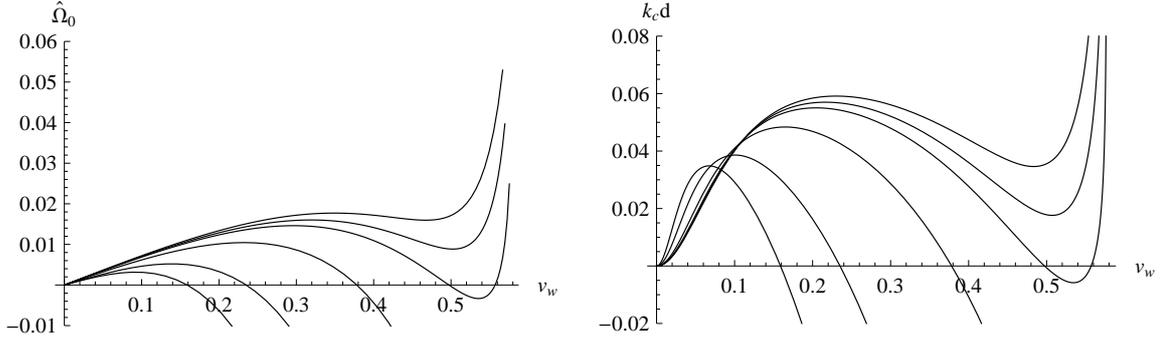}
\caption{The values of $\W_{0}$ and $k_cd$ as functions of the wall velocity,
for $\baL=0.1$ and,
from bottom to top,  $T_+/T_c=0.98$, $0.96$, $0.92$, $0.9$, $0.895$ and $0.89$.}
\label{figommaxkcamas}
\end{figure}

The dependence on the amount of supercooling is more important (see
Fig. \ref{figommaxkcamas}). For small supercooling we have
$v_{\mathrm{crit}}\simeq v_c \simeq\sqrt{1-T_+/T_c}$. As we increase
the amount of supercooling, we observe that $v_{\mathrm{crit}}$ grows
more quickly than $v_c$, as predicted in the previous section. Thus,
e.g., for $T_+/T_c=0.92$, we have $v_c\simeq 0.28$ while
$v_{\mathrm{crit}}\simeq 0.38$. We see that this behavior becomes
critical for a value of $T_+/T_c\simeq 0.9$, which corresponds to
$v_c\simeq 0.31$, while the critical velocity reaches a value
$v_{\mathrm{crit}}\simeq 0.5$. Here, a second critical velocity
appears, beyond which $\W_{0}$ and $k_c$ become positive again.
Increasing slightly the amount of supercooling, the critical velocity
ceases to exist and the deflagration is unstable in all the range
$0<v_w<c_s$.

Notice that, as we increase the supercooling, the non-relativistic
regime of $v_w$ does not suffer qualitative modifications, while the
relativistic regime changes considerably. This happens because  the
difference $\Delta v=v_+-v_-$ grows  with $v_w$. As discussed in Sec.
\ref{arbv}, a high value of $\Delta v$ prevents $\W_{0}$ and $k_c$ to
become negative. This effect is so important that, beyond a certain
amount of supercooling,  $v_{\mathrm{crit}}$ disappears before
reaching $c_s$. Moreover, the value of $\W_{0}$ at $v_w=c_s$ begins
to grow very quickly, and the value of $k_c$ diverges.

\subsection{Reheating effects}

As already discussed, it is important to notice that $T_+$ is not the
nucleation temperature, and should not be considered as a free
parameter. The reheating in front of the wall increases with the wall
velocity. Hence, if we fix $T_N$ and increase $v_w$, the temperature
$T_+$ will get closer to $T_c$, reducing the value of $v_c$.

\begin{figure}[bth]
\centering
\epsfysize=7cm \leavevmode \epsfbox{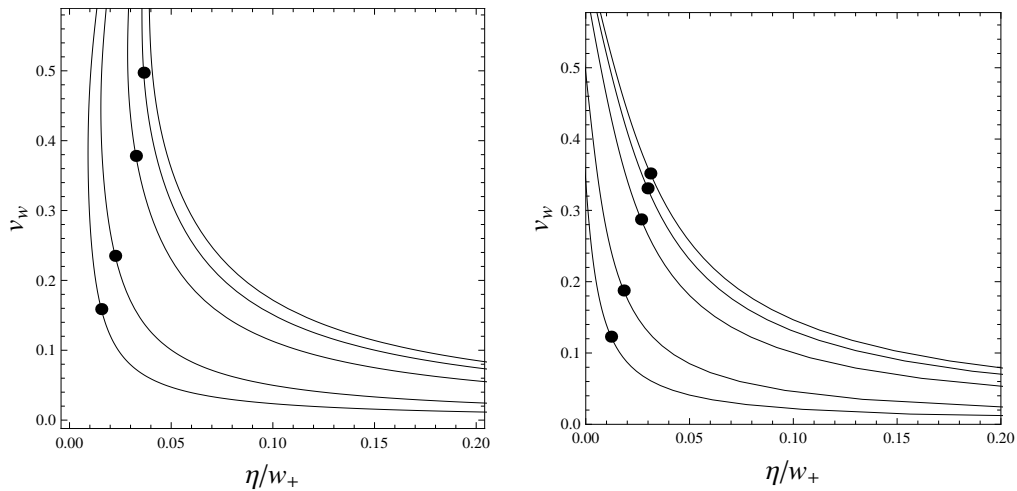}
\caption{The wall velocity as a function of the friction,
for $\baL=0.1$. In the left panel $T_+$ is fixed. In the right panel,
$T_N$ is fixed. The values of $T_+/T_c$ and $T_N/T_c$ are,
from bottom to top,  $0.98$, $0.96$, $0.92$, $0.9$ and $0.89$.}
\label{figvelocity}
\end{figure}

To see the importance of this effect, let us consider the wall
velocity as a function of the friction coefficient $\eta$. In the
left panel of Fig. \ref{figvelocity} we show the relation between
velocity and friction for fixed $T_+$, corresponding to some of the
curves of Fig. \ref{figommaxkcamas}. The dots indicate the critical
velocity. Thus, below the dots the deflagration is unstable under
long wavelength perturbations ($\lambda>1/k_c$). For the upper curve,
the critical velocity does not exist and deflagrations of any
velocity have instabilities. Notice that (fixing $T_+$) deflagrations
do not exist for small enough $\eta$. Besides, for some values of the
friction, there are two possible wall velocities.

In the right panel of Fig. \ref{figvelocity}, we fix instead the
value of $T_N$ (to the same values given previously to $T_+$). For
small $v_w$ (large $\eta$), the results are similar, indicating that
$T_+\simeq T_N$. However, for higher $v_w$ (smaller $\eta$) it
becomes apparent that the velocity is smaller than in the left panel.
This is because $T_+$ is closer to $T_c$ and hydrodynamics acts as an
effective friction. In particular, for small  supercooling (lower
curves), we see that the deflagration is always subsonic, even for
$\eta=0$.  This  means that, depending on the parameters, not any
velocity will be physically reachable. This fact  may be missed when
we consider $v_w$ instead of $\eta$ as a free parameter.

\begin{figure}[bth]
\centering
\epsfysize=5cm \leavevmode \epsfbox{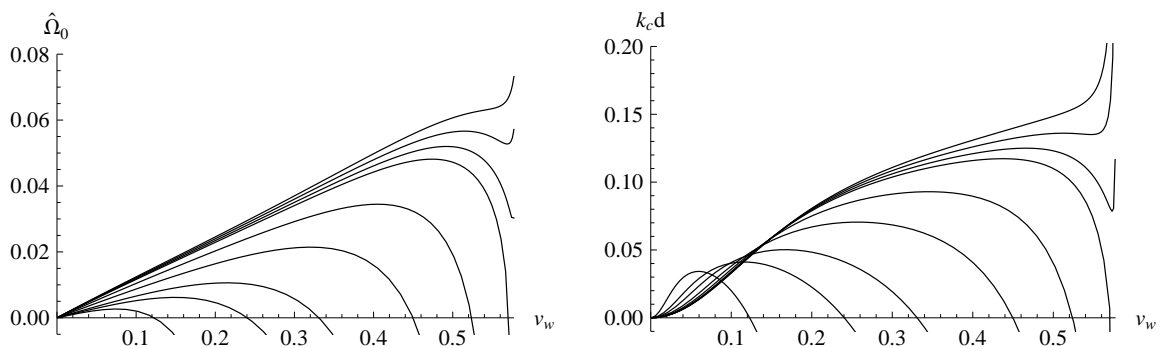}
\caption{The values of $\W_{0}$ and $k_cd$ as functions of the wall velocity,
for $\baL=0.1$ and different values of the nucleation temperature.
From bottom to top, we have $T_N/T_c=0.98$, $0.94$, $0.9$, $0.84$, $0.8$,
$0.775$, $0.77$, $0.765$
and $0.76$.}
\label{figommaxkcan}
\end{figure}

In Fig. \ref{figommaxkcan} we plot again the values of $\W_{0}$ and
$k_c$, this time fixing $T_N$ and taking into account the reheating.
We considered some of the previous values of $T_N$, as well as higher
amounts of supercooling. We see that the behavior is softened with
respect to Fig. \ref{figommaxkcamas}. The critical velocity now grows
more slowly, and reaches $v_{\mathrm{crit}}=c_s$ at a temperature
$T_N\simeq 0.775$. As the amount of supercooling is increased
further, wall velocities close to the speed of sound become more and
more unstable.

Thus, we may have two different situations, depending on the amount
of supercooling. If the latter is small enough, we have
$v_{\mathrm{crit}}<c_s$ and the deflagration is stable in the range
$v_{\mathrm{crit}}<v_w<c_s$. Since the weak deflagration matches the
supersonic Jouguet deflagration at $v_w=c_s$, this may be an
indication of the stability of the latter. On the other hand, after
the critical velocity reaches the value $c_s$, the situation is
inverted; velocities close to $c_s$ are the most unstable ones. This
instability may be an indication of the instability of the supersonic
Jouguet deflagration. This result suggests that, under the conditions
which give weak deflagration velocities close to $c_s$ (a high amount
of supercooling and a low friction), another solution is
hydrodynamically favored, namely, the weak detonation.

\section{Effects of the instability} \label{conseq}

In this section we shall consider the effects of the instability on
the dynamics of a cosmological phase transition.

\subsection{Bubble growth and surface corrugation}

As mentioned in Sec. \ref{scales}, bubbles nucleate with an initial
radius which is of the order of the scale $d$. Their walls accelerate
during a time which is also of order $d$, after which they reach a
terminal velocity. The scale $d$ is in general much smaller than the
final bubble radius or the duration of the phase transition (the
latter two are related by $R_f\sim v_w\Delta t$). Indeed, although
both $d$ and $\Delta t$ depend on the non-trivial dynamics of the
phase transition, the former is determined by forces which are not
related to the expansion rate of the universe, $H$, whereas the
latter will be a fraction of the age of the Universe, $t\sim H^{-1}$.
Roughly, we have $d\sim T^{-1}$ and $\Delta t\sim M_P/T^2$, where
$M_P$ is the Planck mass. Hence, we have $d/\Delta t\sim T/M_P$,
which is, for most phase transitions, many orders of magnitude less
than 1. Therefore, the terminal velocity is reached almost
immediately. During most of its growth, the bubble will be in a
stationary state, unless the growth becomes unstable under small
perturbations\footnote{In Ref. \cite{hkllm} the results of the
stability analysis were applied to the acceleration stage (although
they were derived for the stationary motion). The conclusion was
that, since the terminal velocity is reached when the bubble size is
$\sim d<\lambda_c$, the growth is not destabilized during this stage.
We shall assume that the wall reaches the stationary state before it
can become unstable.}.

Let us assume that we have instabilities, and consider their growth.
We shall concentrate on exponentially unstable perturbations (i.e.,
$\Omega>0$). We remark, though, that there is a velocity $v_c'$
around which perturbations of any wavelength are marginally unstable
(see the discussion around Figs. \ref{figlinear} and
\ref{figkcommaxnr}).

In most cases, the behavior of the instability can be described (at
least qualitatively) by the analytic approximations derived Sec.
\ref{stab}, except in the limit in which both $v_{\mathrm{crit}}$ and
$v_w$ are very close to $c_s$, where the behavior departs
significantly from these approximations. For $v_w<v_{\mathrm{crit}}$,
the essential features of the instability  are already present in
Link's result, namely, that $\Omega$ is of order $\baL$ and
proportional to the wall velocity, as well as the dependence
$\Omega(k)\propto k(k-k_c)$ [see Eq. (\ref{omegalink})]. Thus, for
the purposes of the present discussion, we shall use the
approximations (\ref{kcdlinear}-\ref{omegamaxlinear}), to lowest
order in $\baL$. In particular, we have
 \beg
  k_cd\simeq \fr{\beta{\baL}/{2}}{\beta+{\baL}/{2}}(1-\beta).
 \en
In most cases the parameter $\baL$ will be small. Hence, for
$\beta\sim 1$ we have
 \beg \label{kcbeta1}
  k_cd\simeq \fr{\baL}{2} (1-\beta) \quad \quad (\beta\sim 1).
 \en
However, in some cases we may have a very small velocity (e.g., due to a
significant reheating during bubble expansion). In such a case we have
 \beg \label{kcbetall1}
  k_cd\simeq  \beta \quad \quad (\beta\ll 1).
 \en
In any case, we may neglect $kd$ in the denominator of Eq.
(\ref{omegalinear}) and write
 \beg
  {\Omega}/{k} \simeq \W_{0}(1-k/k_c),\label{omaprox}
 \en
with
 \beg
 \W_0\simeq \baL v_w(1-\beta)/2.
 \en

\begin{figure}[bth]
\centering
\epsfysize=5cm \leavevmode \epsfbox{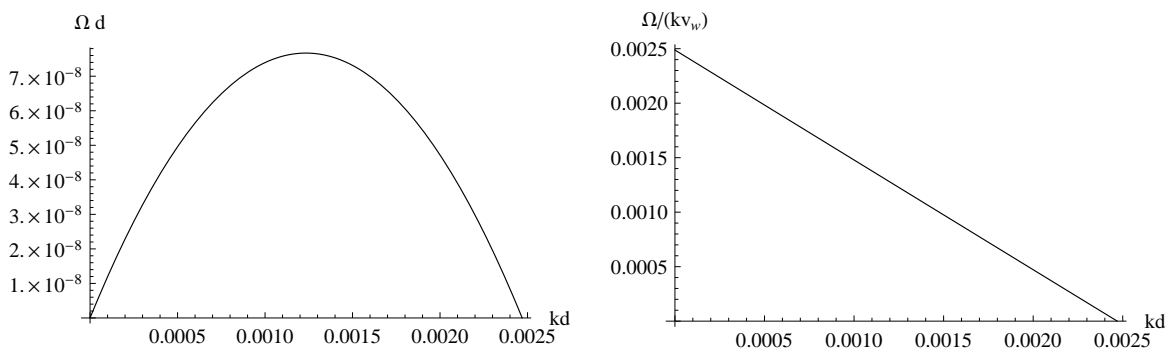}
\caption{The values of $\Omega$ and of the dimensionless quantity
$\Omega/(kv_w)$ as functions of $k$, for
the same parameters of Fig. \ref{figlinear}.
}
\label{figomega}
\end{figure}

Thus, perturbations above the critical wavelength $\lambda_c=1/k_c$
are unstable. Notice, though, that stability is recovered for
$\lambda\to\infty$. Indeed, the finite value of $\W_0$ implies that
$\Omega$ vanishes at $k=0$ (see Fig. \ref{figomega}, left panel). The
stability of the zero mode may be understood as follows. This
perturbation corresponds to acceleration of the wall without
corrugation. However, we know that the uncorrugated wall has already
undergone an acceleration stage and has reached a terminal velocity,
ending in this  stationary state. In a way, the stability for $k=0$
just confirms the existence of such a stationary state. On the other
hand, if we allow the wall to be deformed, instabilities arise. In
this case, the corrugation introduces a length scale, and the
relevant quantity will not be the value of $\Omega$ but the
dimensionless combination $\Omega/k$. The latter is a velocity and in
principle should be contrasted with $v_w$. Thus, an important
parameter will be $\W_0/v_w\sim k_c d\lesssim \baL/2$  (see Fig.
\ref{figomega}, right panel).

As can be seen in the left panel of Fig. \ref{figomega}, there is a
mode with maximum growth rate. For the approximation (\ref{omaprox}),
the wavenumber of this mode is $k=k_c/2$, and the growth rate is
$\Omega_{\max} =\W_0 k_c/4$. We may obtain a stability criterion by
considering this mode, which has the shortest growth time $\tau \sim
\Omega_{\max}^{-1}$ \cite{link,kaj}. Notice that bubbles have  a
finite size $R_b$ and, thus, cannot admit corrugations of arbitrary
scales. At the beginning, bubbles are very small, $R_i\sim d$. From
the equations above, we see that the critical wavelength is higher
than that, $\lambda_c> 2d/\baL$. Hence, physical perturbations (i.e.,
those with $\lambda<R_b$) will be stable until bubbles reach a size
$R_b>\lambda_c$. Besides, the time available for an instability to
grow is bounded by the duration of the phase transition, $\Delta
t\approx R_f/v_w$. The mode with the shortest growth time
$\Omega_{\max}^{-1}$ will be able to develop if $R_f$ is larger than
the corresponding wavelength $2/k_c$, and if $\Delta t>
\Omega_{\max}^{-1}$. The two conditions are thus $R_f>2/k_c$ and
$\Delta t>4/(\W_0k_c)$. The latter condition (which implies the
former) gives
 \beg
  {R_f} >  \fr{4/k_c}{\W_0/v_w}.
  \label{condlink}
 \en

This instability criterion can be improved, if we notice that modes
with longer growth times may have, on the other hand, more time
available to develop. Indeed, a perturbation with wavenumber $k$ can
only be formed after the bubble reaches a size $\lambda=1/k$. The
time elapsed since $\lambda$ ``enters'' the bubble size until the
bubble reaches a larger size $R_b$ is given by $\Delta
t_k=({R}_b-1/k)/v_w$. On the other hand,  the perturbation (if
unstable)  grows in a time $\tau\sim\Omega^{-1}$. Therefore, the mode
$k$ will become dynamically important when $R_b$ is such that
$\Omega\Delta t_k\gtrsim 1$. Using the approximation (\ref{omaprox})
we obtain
 \beg \label{stabcond}
  \Omega\Delta
t_k = \fr{\W_{0}}{v_w}\left(1-\fr{\lambda_c}{\lambda}\right)
  \left(\fr{{R}_b}{\lambda}-1\right).
 \en
This equation takes into account the fact that perturbations are
linearly unstable only in the range $\lambda_c<\lambda<R_b$. The
first two factors in Eq. (\ref{stabcond}) are smaller than $1$.
Moreover, we have in general ${\W_{0}}/{v_w}\ll 1$. However, the last
factor may be large, depending on the final bubble size.

For a given $R_b$, the dynamically most relevant perturbation is now
given by the maximum of $\Omega\Delta t_k$,
 \beg \label{omdynmax}
  (\Omega\Delta t_k)_{\max}=\fr{\W_{0}}{v_w}
  \fr{({R}_bk_c-1)^2}{4k_c{R}_b},
 \en
which is attained for a wavenumber
 \beg
 {k}=\fr{1}{2}\left(\fr{1}{{R}_b}+k_c\right).
 \label{kdynmax}
 \en
This perturbation will be important if $(\Omega\Delta
t_k)_{\max}\gtrsim 1$. If we apply this criterion to the final bubble
size, for which we have $R_f\gg 1/k_c$, we obtain Eq.
(\ref{condlink}). On the other hand, when the bubble size is still
comparable to the critical wavelength, the criterion says that no
instability is important. The instabilities become important once the
bubble reaches the size
 \beg \label{rbinst0}
  {R_b^{\mathrm{inst}}}= \fr{1/k_c}{\W_0/v_w}
  \left[{1+\sqrt{1+\W_0/v_w}}\right]^2 .
 \en
Since in general $\W_0/v_w$ is small, we have
${R_b^{\mathrm{inst}}}\approx (4/k_c)/(\W_0/v_w)$, as in Eq.
(\ref{condlink}).

The parameters $\W_0$ and $k_c$ are not independent. For $\beta\sim
1$ we can use the approximation (\ref{kcbeta1}), which gives
$k_cd\approx \W_{0}/v_w\approx (1-\beta)\baL/2$. Thus, we have
 \beg \label{rbinst}
  {R_b^{\mathrm{inst}}}\simeq
  \left[\fr{4/\baL}{1-({v_w}/{v_{\mathrm{crit}}})^2}\right]^2 d \quad \quad (\beta\sim
  1).
 \en
For $v_w\to v_{\mathrm{crit}}$, Eq. (\ref{rbinst}) diverges, meaning that
the instabilities need infinite time to develop. On the other hand, for
$v_w\to 0$ we must use the approximation (\ref{kcbetall1}). Although
$\W_0/v_w$ does not vanish in this limit, $k_c$ does, and we have
 \beg \label{rbinstbetall1}
  {R_b^{\mathrm{inst}}}\simeq
  \fr{8}{\baL}\fr{v_{\mathrm{crit}}^2d}{v_w^2}\simeq \fr{8}{\baL}
  \fr{\sigma/L}{v_w^2}   \quad \quad (\beta\ll 1),
 \en
where we have used the relations $d=\sigma/F_{\mathrm{dr}}$ and
$F_{\mathrm{dr}}\simeq Lv_c^2$ to make explicit the fact that
$v_{\mathrm{crit}}^2d$ does not depend on the temperature $T_N$. In
most cases, we will have $\beta\sim 1$. Therefore, Eq. (\ref{rbinst})
can be used to determine the bubble size at which the instabilities
become important. Roughly, we have $R_b^{\mathrm{inst}}/d\sim
(4/\baL)^2$ (unless there is a fine tuning so that $v_w\simeq
v_{\mathrm{crit}}$). Thus, $R_b^{\mathrm{inst}}$ will be in general
quite higher than the initial bubble size $d$. On the other hand, as
we have mentioned, the \emph{final} bubble size $R_f$ will be a
fraction of the Hubble radius $H^{-1}$, which is many orders of
magnitude larger than $d$ and, in general, much larger than
$R_b^{\mathrm{inst}}$ as well. Notice also that, as can be seen from
Eq. (\ref{rbinst0}) we have, in general, $R_b^{\mathrm{inst}}\gg
1/k_c$. As a consequence, the dynamically most relevant wavenumber
will be, according to Eq. (\ref{kdynmax}), close to $k\simeq k_c/2$.

As pointed out in Ref. \cite{kaj}, the stability may be recovered due
to reheating. Indeed, once the shock fronts (which precede the phase
transition fronts at a speed $v_{\mathrm{sh}}\simeq c_s$) meet, they
may reheat the space back to a temperature $T_r$ very close to $T_c$.
In such a case, a phase equilibrium stage begins, during which the
regions of stable phase can grow only at the rate at which the
adiabatic expansion takes the latent heat away \cite{w84,h95,ms08}.
From the viewpoint of the instabilities, the effect would be,
roughly, to change the boundary condition for the temperature from
$T_N$ to $T_r$. As a consequence, both velocities $v_c$ and $v_w$
will decrease significantly. It is not clear which will be the value
of $\beta$. Nevertheless, since $d\simeq \sigma/(Lv_c^2)$, we see
that either of Eqs. (\ref{rbinst}) and (\ref{rbinstbetall1}) will
give a large value of $R_b^{\mathrm{inst}}(T_r)$. To appreciate the
importance of the change of the scale $d$ after reheating, we notice
that the new wall velocity is proportional to the expansion rate,
$v_w\sim R_b H/\baL$ \cite{h95,m04}. Thus, Eq. (\ref{rbinstbetall1})
gives an enhancement $\sim (H^{-1}/R_b)^2$, which ensures the
stability of the deflagration. Notice that the time available for the
instabilities to grow is the same as before ($\Delta t\sim R_f$),
since the spacing between bubble centers is given by $R_f$ and the
reheating will occur after a time of order $\Delta
t=R_f/v_\mathrm{sh}$. The only difference is that, if a phase
equilibrium stage is reached, the \emph{total} duration of the phase
transition will be longer than $\Delta t$.

In summary, the time and length scales are the following. After a
bubble nucleates with size $\sim d$, it reaches the stationary motion
in a time $\sim d$. The instabilities (provided that
$v_w<v_{\mathrm{crit}}$) become dynamically relevant much later,
after a time $\sim (4/\baL)^2 d$. In general, though, there will be
ample time for this to happen during bubble expansion, since the
final bubble size is still much larger ($R_f\sim$ a fraction of
$H^{-1}$). After a time $\sim R_f$, the phase transition may end, or
it may enter a phase-equilibrium stage during which stable growth is
recovered. The dynamically relevant unstable modes are those of
wavelengths $\sim 4d/\baL$. This is generally much shorter than the
average bubble size $R_f$ by the end of the phase transition.

\subsection{Deflagrations vs detonations}

For strong supercooling or small friction, the stationary solution is
in general a detonation with high velocity. On the contrary, for
$T_N$ very close to $T_c$ or large $\eta$ we generally have weak
deflagrations with small velocities. Between these two extremes, we
may have coexistence of subsonic and supersonic solutions (see e.g.
\cite{ms09,ms12}). In such a case, the question arises of which one
will be realized during the phase transition, and of whether this can
be elucidated by the analysis of instabilities. These issues were
discussed previously from a different approach, namely, in the
context of a numerical investigation of Eqs.
(\ref{EM1.1}-\ref{EM1.2}) in a grid \cite{ikkl94}. Instead of using
an approximation for the interface, the field configuration
$\phi(\mathbf{x},t)$ was considered together with the fluid profile.
Thus, the dynamical evolution of a phase transition front was
studied, from the initial acceleration period to the collision
between two bubbles.

Regarding the coexistence of deflagrations and detonations, it was
found in \cite{ikkl94} that, in the cases in which either of the
stationary solutions is possible, it is the detonation the one which
is realized during bubble expansion. Nevertheless, this seems to be
due to the dynamical evolution rather than due to an instability of
the deflagration, since the detonation configuration is reached
without going through a deflagration configuration. According to our
results, the instability of the deflagration is unrelated to the
existence of detonation solutions. Although the instabilities become
important for large amounts of supercooling and low friction,
detonations already exist for more moderate values of these
parameters, where deflagrations are still stable.

Regarding possible instabilities, the results of Ref. \cite{ikkl94}
were the following. The wall configuration was found to be unstable
only for strong deflagrations and for detonation solutions close to
the Jouguet point. In contrast, if given as an initial condition, the
weak deflagration remains as such, indicating stability. This seeming
contradiction with the instability of deflagrations found in the
present paper and in Ref. \cite{hkllm} has a simple explanation. Due
to calculation convenience, the amounts of time considered in
\cite{ikkl94} (as well as the distance between bubbles) were much
less than those in an actual cosmological phase transition. This is a
general problem of lattice calculations. A similar calculation was
carried out recently \cite{hhrw13}. In this case, the available time
did not even allow the walls to reach the stationary state before
they collided. As we have seen, the time needed for the instability
to become dynamically relevant is much longer than the time it takes
to reach the stationary state. Nevertheless, the duration of the
phase transition is still much longer.

\section{A physical model}
\label{model}

The strength of the phase transition \cite{quiros} depends on the
separation between the two minima of the free energy, $\Delta
\phi=\phi_+-\phi_-$, and is usually characterized by the value of
$\Delta \phi/T_c$ (for instance, in the limit $\Delta\phi\to 0$ one
gets a second order phase transition). However, the phase transition
dynamics does not depend on this single parameter alone. The wall
velocity depends essentially on three parameters. These are the
amount of supercooling (which determines the pressure difference),
the latent heat (which reheats the plasma slowing down the wall), and
the friction coefficient. As can be seen, e.g., in Eq.
(\ref{rbinst}), these parameters are relevant for the dynamics of the
instabilities as well. The ratio $T_N/T_c$ determines the value of
$v_c$ and, hence, of $\beta$. The ratio $L/w_+$ gives the parameter
$\baL$. Finally, the friction coefficient determines the wall
velocity $v_w$.

Unfortunately, these parameters do not have a simple relation in
general\footnote{While the latent heat can be directly computed from
the free energy density $\mathcal{F}(\phi,T)$, the calculation of
$T_N$ involves, first, calculating the nucleation rate (using thermal
instantons \cite{linde}) and, then, considering the dynamics of the
phase transition to compute the number of bubbles nucleated in a
causal volume \cite{gw81}.}. For a given model, both the released
energy and the amount of supercooling increase with the strength of
the phase transition. Indeed, a higher value of $\Delta \phi$ implies
a higher discontinuity of the energy density (i.e., a higher $L$) as
well as a wider and higher barrier between minima. The latter causes
the system to stay longer in the metastable minimum, i.e., a lower
temperature will be reached before bubble nucleation effectively
begins. A strong supercooling causes a large pressure difference
between phases and, thus, favors a high wall velocity. In contrast, a
large release of latent heat causes reheating and slows the wall
down. Besides, the wall velocity depends on the friction coefficient.
This parameter is quite difficult to calculate, model-dependent, and
is the main source of uncertainty for the wall velocity\footnote{In
general, it depends on the couplings of the particles with the field
$\phi$ (the stronger the coupling, the higher the friction). However,
it also depends on the particles interactions which determine the
diffusion of particle densities near the wall \cite{micro}.}.
Therefore, it is not easy to ascertain, without a detailed
calculation, whether the wall velocity will be above
$v_{\mathrm{crit}}$ or not.

We shall address elsewhere such a detailed study of specific models.
Here we wish to discuss in general the possibility that the
deflagration becomes unstable in physical models, as well as some
possible cosmological implications of the instability. For that aim,
we shall consider the case of the electroweak phase transition, which
may be quite different for different extensions of the Standard
Model.

\subsection{The electroweak phase transition}

In Ref. \cite{hkllm}, the electroweak phase transition was considered
for the minimal Standard Model (SM), with a Higgs mass $m_H=40 GeV$
in order to obtain a first-order phase transition which is strong
enough to fulfil the requirement of  electroweak baryogenesis
($\Delta\phi/T_c\gtrsim 1$). Still, the phase transition for such a
model is relatively weak and has a small amount of supercooling. The
critical velocity was found to be bounded by $0.07$, whereas
microscopic calculations gave $v_w\gtrsim 0.1$ \cite{dlhll}. It was
thus concluded that the propagation of the phase transition front as
a deflagration is stable.

For the actual value of the Higgs mass, the SM electroweak phase
transition is just a smooth crossover. Nevertheless, many extensions
of the SM have been considered in the literature. In particular, the
Minimal Supersymmetric Standard Model (MSSM) has been extensively
investigated in relation with electroweak baryogenesis (see, e.g.,
\cite{mssm}). Moreover, it is well known that extra scalar singlets
may cause an extremely strong phase transition (see, e.g.,
\cite{scalars}). Thus, depending on the model (and on the model
parameters) the electroweak bubble may grow either as a deflagration
or as a detonation (see, e.g., \cite{ms10}), or it may even run away
\cite{bm09}. The former possibility favors baryogenesis, whereas the
latter two favor GW generation.

\begin{figure}[bth]
\centering
\epsfysize=7cm \leavevmode \epsfbox{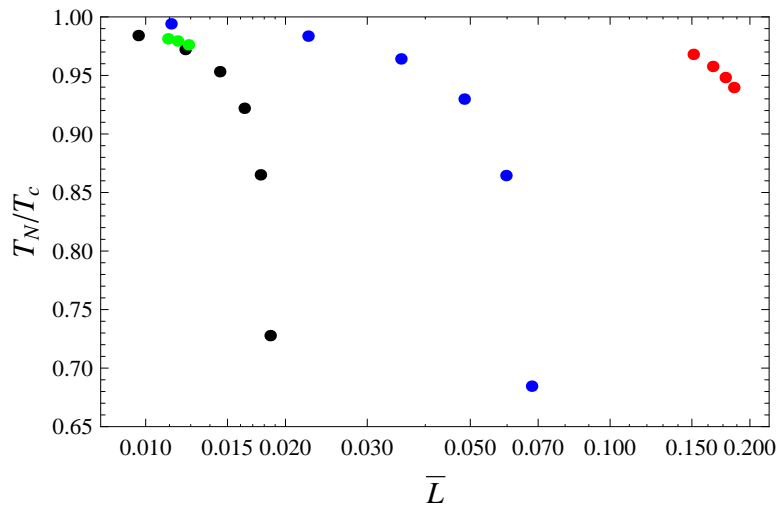}
\caption{\small The space of the parameters $\baL$ and $T_N/T_c$ for the
electroweak phase transition,
for some extensions of the Standard Model (taken from Ref. \cite{lms12}).
Black dots correspond to extra scalar singlets,
with $g_s=12$ degrees of freedom. The strength of the phase transition
increases with the coupling of the scalars to the Higgs, $h_s$.
Thus, a higher $h_s$ implies a larger latent
heat as well as a smaller $T_N$. The
values of $h_s$ range from $0.7$ to $1.2$ and are equally spaced.
Blue dots correspond to the same extension, with less singlets,
$g_s=2$. In this case we have $1.4\leq h_s\leq 1.9$. Red dots corresponds to
an extension with heavy fermions and bosons \cite{cmqw05}. The
coupling of the fermions to the Higgs is in the range $2.2\leq h_f\leq 2.8$
(the strength of the phase transition increases with $h_f$). Green
dots corrrespond to the MSSM in the light stop scenario, for
stop masses $m_{\mathrm{stop}}=132,136$, and $140$GeV (from right to left).}
\label{figmodels}
\end{figure}

In Fig. \ref{figmodels} we show the values of $L$ and $T_N$ for some
extensions of the Standard Model.
We only considered phase transitions with $\Delta \phi/T>1$. Weaker
phase transitions give less supercooling and smaller latent heat (in
the limit of a second order phase transition we have $T_N=T_c$ and
$L=0$). Thus, the models considered in Fig. \ref{figmodels} give
values of the latent heat in the range $\baL\sim 10^{-2}-10^{-1}$.
Higher values of $\baL$ may be possible in other models (the physical
bound is $\baL<1$).
On the other hand, it may be inferred from Fig. \ref{figmodels} that
a very strong supercooling is hard to achieve in a physical model.
Indeed, the two lower dots (those around $T_N/T_c=0.7$) were obtained
for  phase transitions which are already extremely strong (a slightly
stronger phase transition would remain stuck in the false vacuum,
causing an inflationary era). The numerical examples considered in
Sec. \ref{numres} roughly spanned this region of parameter space.

As we have seen, the possible instabilities need time to grow. The
initial bubble radius at nucleation, which is related to the size
scale $d$, is in general $R_i\sim 10 /T$. On the other hand, the
instabilities become dynamically important for bubble sizes larger
than $R_b^{\mathrm{inst}}\sim (4/\baL)^2d$, which is thus in the
range $\sim 10^4/T-10^6/T$. The average bubble size by the end of the
phase transition is a fraction of the Hubble size $H^{-1}$. The exact
value depends on the whole dynamics of the phase transition, and is
not easy to estimate without a complete numerical calculation.
Numerical results (see, e.g., \cite{ma05,ms08}) give values which
range from $R_b\sim H^{-1}$ for very strong phase transitions to
$R_b\sim 10^{-5} H^{-1}$ or smaller for weak phase transitions. At
the electroweak scale we have $H^{-1}\sim M_P/T^2$, with $T\sim
100\mathrm{GeV}$ and $M_P\sim 10^{19}\mathrm{GeV}$. Hence, the final
bubble size for the electroweak phase transition will be in the range
$R_f\sim 10^{12}/T-10^{17}/T$. Thus, this example confirms the
general hierarchy $R_i\ll R_b^{\mathrm{inst}}\ll R_f$ obtained in the
previous section. The most important perturbations will be those with
$\lambda\sim 4d/\baL \sim 10$-$10^2 d\gtrsim R_i$.

\subsection{Baryogenesis and gravitational waves}

This phase transition may have several cosmological consequences,
most of them depending on the dynamics of moving walls. In principle,
the hydrodynamic instability  may affect any of the cosmological
remnants. For instance, the generation of magnetic fields due to
instabilities of the bubble walls was considered in Ref.
\cite{soj97}. Here we wish to discuss the generation of gravitational
waves (GW) and of the baryon asymmetry of the universe (BAU), which
require quite different values of the wall velocity.

A successful electroweak baryogenesis requires $\Delta\phi/T_c\gtrsim 1$, so
that baryon number violating processes (sphalerons) are turned off in the
broken-symmetry phase, in order to avoid the washout of the generated BAU.
Regarding the wall velocity, on the one hand, it should not be too large, so
that sphalerons have enough time to generate baryons in the symmetric phase
(sourced by $CP$-violating interactions of the wall with the particles of
the plasma). On the other hand, the wall velocity should not be too small
either, in order to avoid that sphalerons in the symmetric phase have enough
time to reach the equilibrium and wash out the generated BAU. All in all, a
relatively small wall velocity ($v_w\sim 10^{-2}-10^{-1}$) is needed. As a
consequence, baryogenesis is favored for relatively weak phase transitions,
which may give such small wall velocities\footnote{The fact that the
incoming flow velocity $|v_+|$ is smaller than $v_w$ may increase the upper
bound \cite{n11}. Moreover, the possibility of electroweak baryogenesis with
detonations has been recently discussed \cite{cn12}.}. Weak phase
transitions will generally have little supercooling and, consequently, small
values of $v_{\mathrm{crit}}$ as well. Therefore, the presence of
hydrodynamic instabilities will depend on details of the specific model.

To see the effect of these potential instabilities on electroweak
baryogenesis, let us assume that $v_w$ is below the critical velocity
$v_{\mathrm{crit}}$. As we have seen, for a weak phase transition,
the instability will become dynamically important when the bubble
reaches a size $R_b^{\mathrm{inst}}\gtrsim 10^6/T$. After that
moment, the growth of the bubble may be of dendritic type
\cite{fa90}. One expects that the motion of the wall will become too
quick to successfully generate baryons \cite{kf92,a97}. We may thus
assume that baryogenesis stops as soon as bubbles reach the size
$R_b^{\mathrm{inst}}$. Since the final bubble size is $R_f\sim
10^{6}R_b^{\mathrm{inst}}$, we see that the resulting BAU will be
strongly suppressed with respect to a stable wall\footnote{In this
argument we have used the rough approximation
$R_b^{\mathrm{inst}}/d\sim (4/\baL)^2$ for $v_w<v_{\mathrm{crit}}$.
Taking into account the factor $1/(1-v_w^2/v^2_{\mathrm{crit}})$ will
not change the conclusion, unless $v_w$ is very close to
$v_{\mathrm{crit}}$.}. We see that an accurate determination of the
wall velocity becomes crucial since electroweak baryogenesis may be
completely spoiled if $v_w<v_{\mathrm{crit}}$.

On the other hand, we have seen that, once shock fronts meet and
reheat the plasma, the motion of phase transition interfaces as
stable deflagrations may be reestablished. Depending on the friction
and latent heat, the value of the wall velocity during this
phase-equilibrium stage may or may not be  appropriate  for
baryogenesis \cite{h95,ma05,m01}. In case it is, bubble walls will
generate baryons during the last stages of the phase transition. It
is important to notice that a significant fraction of space may be
spanned by the walls during this stage.

Generating gravitational waves of sizeable intensity generally
requires quite higher velocities ($v_w> c_s$) in order to generate a
strong disturbance of the plasma (through bubble collisions and
turbulence). Hence, the instability of the deflagration is
preferable, as it accelerates the wall motion. In fact, gravitational
waves of sizeable amplitude seem to be possible only in models with
large amounts of supercooling (e.g., the lower dots in Fig.
\ref{figmodels}), which give detonations with high velocities
\cite{lms12}. Such models may also allow deflagrations with
velocities $v_w$ close to $c_s$ or higher.  In general, these models
will give $v_{\mathrm{crit}}$ also close to $c_s$. According to Fig.
\ref{figommaxkcan}, in this case the deflagration may have
instabilities on all wavelengths (notice the divergence of $k_c$ at
$v_w= c_s$). This opens the possibility of a new mechanism of GW
generation, which may compete with the collisions of detonations,
even for weaker phase transitions.

The evolution of the system beyond the linear regime is difficult to
guess. Furthermore, it will be characterized by turbulent motions of
the fluid, which make the treatment more involved. The results of a
simple geometrical model (described by an equation which depends only
on the local geometry of the interface) suggest that the growth may
be of dendritic type \cite{fa90}. This means that ``fingers'' grow
out of the wall and then split into new fingers.

A spherically symmetric bubble cannot generate gravitational
radiation. As a consequence, the usual mechanisms (bubble collisions
and turbulence) rely on the collision of bubble walls, once bubbles
have grown up to there final size. The corrugation instability, in
contrast, deforms the walls and stirs the fluid as soon as the bubble
reaches the size $R_b^{\mathrm{inst}}\sim (4/\baL)^2R_i$, when
bubbles are still much smaller than the final mean size. Therefore,
the GW spectrum will be quite different. The characteristic
wavelength of the gravitational radiation is given by the stirring
scale. For the usual mechanisms, this is roughly the bubble size
scale $R_b$, which is determined by the mean average separation
between nucleation points. In the case of unstable growth, the
relevant scale (or scales) will be smaller.

Initially, the source of turbulence will be the unstable corrugations
of the wall (accompanied by perturbations of the fluid). Thus, the
initial stirring length scale is that of the most relevant unstable
mode, $\lambda_{\mathrm{inst}}\sim 2/k_c\sim 4R_i/\baL$. These
perturbations then grow in size and amplitude. In the case of
dendritic growth, a new length scale may arise, namely, the length of
the fingers. In any case, after a certain time the turbulent fluid
will ``see'' also the nominal radius of the bubbles $R_b\sim
R_b^{\mathrm{inst}}$. This gives another stirring scale. The bubble
spacing $R_f$ may also play a role in the turbulence spectrum. As we
have seen, both $\lambda_{\mathrm{inst}}$ and $R_b^{\mathrm{inst}}$
are much smaller than $R_f$. For the usual mechanisms, the
(redshifted) peak of the spectrum is around the miliHertz
(corresponding to $R_f\sim 10^{-2} H^{-1}$). For the unstable
growing, the GW spectrum may have several peaks, some of them at
frequencies much higher than that.

\section{Conclusions} \label{conclu}

The possibility that an observable background of gravitational waves
was produced at the electroweak phase transition has motivated in the
last years a renewed interest in the hydrodynamics associated to the
propagation of phase transition fronts. It is well known that, while
electroweak baryogenesis requires  weak deflagrations with rather
small interface velocities, $v_w\lesssim 0.1$, GW generation is
favored by detonations or runaway solutions with ultra-relativistic
velocities. Thus, the various extensions of the SM give quite
different results, depending on the values of three relevant
parameters, namely, the amount of supercooling, the latent heat, and
the friction. In particular, small supercooling, large friction, and
large latent heat will give in general small wall velocities,
favoring baryogenesis. The  instability of deflagrations may alter
completely this picture.

In this work, we have studied the hydrodynamic stability of
deflagrations. We have calculated the linear instability under
corrugation of the wall as a function of the relevant parameters, we
have analyzed the dynamical relevance of the instabilities, and we
have discussed the implications for the electroweak phase transition
and its cosmological consequences.

The instability of deflagration phase-transition fronts was previously
considered in Ref. \cite{hkllm}. The treatment of that work improved
significantly upon preceding analysis, by taking into account the
perturbations of the force which drives the wall motion. This is an
important aspect, since the pressure difference between phases is very
sensitive to temperature variations. Unfortunately, some simplifications
used for the driving force constrain the application of those results. Our
approach improved several aspects of the calculation of Ref. \cite{hkllm}.
In the first place, we have derived the equation for the perturbations of
the wall directly from the field equation (\ref{EM1.2}), taking into account
\emph{independent} perturbations of the fluid variables on either side of
the wall. This is the main difference with the treatment of Ref.
\cite{hkllm}. Its quantitative effect increases with the wall velocity. We
have also performed a more exhaustive search of instabilities. In
particular, we have looked for complex solutions of the equation for the
exponential growth rate $\Omega$. In the case of a classical burning gas
\cite{landau}, the unstable modes have $\mathrm{Im}(\Omega)=0$. Thus, the
disturbances are not propagated but are only amplified. This feature was
also found (numerically) in the work of Link \cite{link}. We investigated
analytically as well as numerically this possibility for the case of a phase
transition front. The result is that, indeed, we have
$\mathrm{Im}(\Omega)=0$ for $\mathrm{Re}(\Omega)>0$.

For small velocities and small supercooling, our results are qualitatively
similar to those of Refs. \cite{link} and \cite{hkllm}. However, we have
found a range of marginally unstable wavenumbers, which was not noticed in
previous works. Outside this interval we have exponential (either growing or
decaying) behavior. This wavenumber gap arises as a discontinuity at $\Omega
=0$, and is due to the fact that the special mode $q_1(\Omega)$ jumps from
one side of the wall to the other as $\Omega$ changes sign. Studying the
stability in this range would require to go beyond linear perturbations.
Unfortunately, the numerical analysis of Ref. \cite{fa03} did not explore
regions of parameters where our results would differ from those of Ref.
\cite{hkllm}. Moreover, a numerical investigation of the parameter region
where linear perturbation theory predicts instabilities is still lacking.

The general behavior of the linear stability is essentially the following.
Below a critical velocity $v_{\mathrm{crit}}$, perturbations on wavenumbers
$k$ smaller than a value $k_c$ are exponentially unstable. In general, we
have $k_c\lesssim \baL/d$, and $\Omega\lesssim v_w \baL k_c$. The critical
velocity depends strongly on the amount of supercooling. For small
supercooling, we have $v_{\mathrm{crit}}\simeq \sqrt{1-T_+/T_c}$, in
agreement with Ref. \cite{hkllm}. However, as we increase the amount of
supercooling $v_{\mathrm{crit}}$ quickly departs from this simple behavior.
Even taking into account the reheating effect $T_+>T_N$, the critical
velocity soon approaches the speed of sound, which means that any subsonic
velocity becomes unstable. Furthermore, in this case, those velocities which
are closer to the speed of sound have a larger range of unstable wavenumbers
and higher growth rates. This result is in disagreement with  Ref.
\cite{hkllm}, according to which weak deflagrations are always stable in the
limit $v_w\to c_s$. The discrepancy is due to our more realistic treatment
of the equation for the interface.

We have briefly discussed supersonic deflagrations. The case of
supersonic Jouguet deflagrations turns out to be considerably more
involved, and shall be addressed elsewhere. Regarding strong
deflagrations, we have checked, for the case of planar relativistic
phase-transition fronts, that these are trivially unstable, by
showing explicitly that the whole family of  strong deflagrations
(sketched in Fig. \ref{figstrong}, left panel) is not evolutionary.

We have also studied the dynamical importance of the instabilities. Thus, we
have improved the discussions of Refs. \cite{link,kaj}, and we have
established a hierarchy of time and length scales for the growth of bubbles
and instabilities.

We also discussed briefly a physical model, namely, the electroweak phase
transition, and considered two of its possible outcomes, namely, the BAU of
the universe and a stochastic background of gravitational waves. In general,
for a cosmological phase transition, the instabilities have ample time to
develop, provided that $v_w<v_{\mathrm{crit}}$. This may be a serious
problem for electroweak baryogenesis and deserves further investigation for
specific models. On the other hand, the deflagration instability favors the
production of gravitational waves, by accelerating and deforming the walls
almost from the beginning of bubble growth. However, to estimate the GW
spectrum would require to go beyond the linear stability analysis.

\section*{Acknowledgements}

This work was supported by Universidad Nacional de Mar del Plata,
Argentina, grant EXA 607/12.


\begin{thebibliography}{99}

\bibitem{gr01} For a review, see
  D.~Grasso and H.~R.~Rubinstein,
  Phys.\ Rept.\  {\bf 348}, 163 (2001)
  [arXiv:astro-ph/0009061].

\bibitem{vs94} A. Vilenkin and E.P.S.
    Shellard, {\it Cosmic Strings and Other Topological
    Defects} (Cambridge University Press, Cambridge, England,
    1994);
     A.~Vilenkin,
 Phys.\ Rept.\  {\bf 121}, 263 (1985).


\bibitem{w84}
E.~Witten, 
Phys.\ Rev.\ D \textbf{30}, 272 (1984);
G.~M.~Fuller, G.~J.~Mathews and C.~R.~Alcock,
Phys.\ Rev.\ D \textbf{37}, 1380 (1988);
J.~H.~Applegate and C.~J.~Hogan,
Phys.\ Rev.\ D \textbf{31}, 3037 (1985);
H.~Kurki-Suonio,
Phys.\ Rev.\ D \textbf{37}, 2104 (1988);
J.~Ignatius, K.~Kajantie, H.~Kurki-Suonio and M.~Laine,
Phys.\ Rev.\ D \textbf{50}, 3738 (1994) [arXiv:hep-ph/9405336].


\bibitem{h95} A.~F.~Heckler,
Phys.\ Rev.\ D \textbf{51} (1995) 405 [arXiv:astro-ph/9407064];

\bibitem{ma05} A.~Megevand and F.~Astorga,
  Phys.\ Rev.\ D {\bf 71}, 023502 (2005)
  [hep-ph/0409321].

\bibitem{gw} See, e.g., A.~Kosowsky and M.~S.~Turner,
Phys.\ Rev.\ D \textbf{47}, 4372 (1993);
M.~Kamionkowski, A.~Kosowsky and M.~S.~Turner,
Phys.\ Rev.\ D \textbf{49}, 2837 (1994); 
A.~Kosowsky, A.~Mack and T.~Kahniashvili,
Phys.\ Rev.\ D \textbf{66}, 024030 (2002); 
 A.~D.~Dolgov, D.~Grasso and A.~Nicolis,
Phys.\ Rev.\ D \textbf{66}, 103505 (2002); 
 C.~Caprini and R.~Durrer,
Phys.\ Rev.\ D \textbf{74}, 063521 (2006); 
 C.~Caprini, R.~Durrer and G.~Servant,
Phys.\ Rev.\ D \textbf{77}, 124015 (2008) [arXiv:0711.2593 [astro-ph]];
 R.~Apreda, M.~Maggiore, A.~Nicolis and A.~Riotto,
Nucl.\ Phys.\ B \textbf{631}, 342 (2002); 
A.~Nicolis,
Class.\ Quant.\ Grav.\ \textbf{21}, L27 (2004); 
C.~Grojean and G.~Servant,
Phys.\ Rev.\ D \textbf{75}, 043507 (2007); 
 S.~J.~Huber and T.~Konstandin,
  JCAP {\bf 0809}, 022 (2008)
  [arXiv:0806.1828 [hep-ph]];
      S.~J.~Huber and T.~Konstandin,
  JCAP {\bf 0805}, 017 (2008)
  [arXiv:0709.2091 [hep-ph]].
 A.~Megevand,
  Phys.\ Rev.\  D {\bf 78} (2008) 084003
  [arXiv:0804.0391 [astro-ph]];
      J.~Kehayias and S.~Profumo,
  JCAP {\bf 1003}, 003 (2010)
  [arXiv:0911.0687 [hep-ph]].

\bibitem{lms12}     L.~Leitao, A.~Megevand and A.~D.~Sanchez,
  JCAP {\bf 1210}, 024 (2012)
  [arXiv:1205.3070 [astro-ph.CO]].


\bibitem{ckn93} For reviews, see A.~G.~Cohen, D.~B.~Kaplan and
    A.~E.~Nelson,
Ann.\ Rev.\ Nucl.\ Part.\ Sci.\ \textbf{43}, 27 (1993)
[arXiv:hep-ph/9302210]; 
A.~Riotto and M.~Trodden, 
Ann.\ Rev.\ Nucl.\ Part.\ Sci.\ \textbf{49}, 35 (1999)
[arXiv:hep-ph/9901362]. 


\bibitem{dlhll} M.~Dine, R.~G.~Leigh, P.~Y.~Huet, A.~D.~Linde and
D.~A.~Linde, 
Phys.\ Rev.\ D \textbf{46}, 550 (1992) [arXiv:hep-ph/9203203];
 B.~H.~Liu, L.~D.~McLerran and N.~Turok,
Phys.\ Rev.\ D \textbf{46}, 2668 (1992).

\bibitem{micro} See, e.g.,
N.~Turok, 
Phys.\ Rev.\ Lett.\ \textbf{68}, 1803 (1992);
S.~Y.~Khlebnikov,
Phys.\ Rev.\ D \textbf{46}, 3223 (1992); 
P.~Arnold, 
Phys.\ Rev.\ D \textbf{48}, 1539 (1993) [arXiv:hep-ph/9302258];
G.~D.~Moore and T.~Prokopec,
Phys.\ Rev.\ D \textbf{52}, 7182 (1995) [arXiv:hep-ph/9506475];
Phys.\ Rev.\ Lett.\ \textbf{75}, 777 (1995) [arXiv:hep-ph/9503296];
P.~John and M.~G.~Schmidt,
Nucl.\ Phys.\ B \textbf{598}, 291 (2001) [Erratum-ibid.\ B
\textbf{648}, 449
(2003)]; 
G.~D.~Moore,
JHEP \textbf{0003}, 006 (2000). 


\bibitem{hidro} See, e.g., M.~Gyulassy, K.~Kajantie, H.~Kurki-Suonio
    and L.~D.~McLerran,
Nucl.\ Phys.\ B \textbf{237}, 477 (1984); 
H.~Kurki-Suonio,
Nucl.\ Phys.\ B \textbf{255}, 231 (1985); 
K.~Kajantie and H.~Kurki-Suonio,
Phys.\ Rev.\ D \textbf{34}, 1719 (1986); 
K.~Enqvist, J.~Ignatius, K.~Kajantie and K.~Rummukainen,
Phys.\ Rev.\ D \textbf{45}, 3415 (1992). 


\bibitem{landau} L. D. Landau and E. M. Lifshitz, \textit{Fluid
    Mechanics}
(Pergamon Press, New York, 1989).


\bibitem{link}   B.~Link,
  Phys.\ Rev.\ Lett.\  {\bf 68}, 2425 (1992).


\bibitem{hkllm}  P.~Y.~Huet, K.~Kajantie, R.~G.~Leigh, B.~H.~Liu and
    L.~D.~McLerran,
Phys.\ Rev.\ D \textbf{48}, 2477 (1993) [arXiv:hep-ph/9212224].

\bibitem{fa03}  P.~C.~Fragile and P.~Anninos,
  Phys.\ Rev.\ D {\bf 67}, 103010 (2003)
  [gr-qc/0303018].

\bibitem{abney}   M.~Abney,
  Phys.\ Rev.\ D {\bf 49}, 1777 (1994)
  [astro-ph/9305021].

\bibitem{r96}   L.~Rezzolla,
  Phys.\ Rev.\ D {\bf 54}, 1345 (1996)
  [astro-ph/9605033].



\bibitem{gw81} A.~H.~Guth and E.~J.~Weinberg,
Phys.\ Rev.\ D \textbf{23}, 876 (1981). 


\bibitem{ah92} G.~W.~Anderson and L.~J.~Hall,
Phys.\ Rev.\ D \textbf{45}, 2685 (1992). 


\bibitem{m00}   A.~M\'egevand,
  Int.\ J.\ Mod.\ Phys.\ D {\bf 9}, 733 (2000)
  [hep-ph/0006177].

\bibitem{ekns10}
        J.~R.~Espinosa, T.~Konstandin, J.~M.~No and G.~Servant,
        JCAP {\bf 1006}, 028 (2010)
        [arXiv:1004.4187 [hep-ph]];

\bibitem{hs13}
  S.~J.~Huber and M.~Sopena,
  arXiv:1302.1044 [hep-ph].

\bibitem{ariel13}   A.~Megevand,
  JCAP {\bf 1307}, 045 (2013)
  [arXiv:1303.4233 [astro-ph.CO]].

\bibitem{bm09}   D.~Bodeker and G.~D.~Moore,
  JCAP {\bf 0905}, 009 (2009)
  [arXiv:0903.4099 [hep-ph]].

\bibitem{ms12}  A.~Megevand and A.~D.~Sanchez,
  Nucl.\ Phys.\ B {\bf 865}, 217 (2012)
  [arXiv:1206.2339 [astro-ph.CO]].

\bibitem{lm11}
  L.~Leitao and A.~Megevand,
  Nucl.\ Phys.\ B {\bf 844}, 450 (2011)
  [arXiv:1010.2134 [astro-ph.CO]].


\bibitem{ikkl94} J.~Ignatius, K.~Kajantie, H.~Kurki-Suonio and
    M.~Laine,
Phys.\ Rev.\ D \textbf{49}, 3854 (1994); 
  H.~Kurki-Suonio and M.~Laine,
  Phys.\ Rev.\  D {\bf 51}, 5431 (1995)
  [arXiv:hep-ph/9501216];
    H.~Kurki-Suonio and M.~Laine,
  Phys.\ Rev.\ D {\bf 54}, 7163 (1996)
  [hep-ph/9512202].



\bibitem{ms09} A.~Megevand and A.~D.~Sanchez,
Nucl.\ Phys.\ B \textbf{820}, 47 (2009)  [arXiv:0904.1753 [hep-ph]].


\bibitem{quiros}    M.~Quiros,
  arXiv:hep-ph/9901312.


\bibitem{linde}
A.~D.~Linde, 
Nucl.\ Phys.\ B \textbf{216}, 421 (1983) [Erratum-ibid.\ B \textbf{223}, 544
(1983)]; 
Phys.\ Lett.\ B \textbf{100}, 37 (1981). 

\bibitem{stabdeto}   A.~Megevand and F.~A.~Membiela,
  arXiv:1402.5791 [astro-ph.CO].

\bibitem{kn11}  T.~Konstandin and J.~M.~No,
  JCAP {\bf 1102}, 008 (2011)
  [arXiv:1011.3735 [hep-ph]].

\bibitem{s82} P.~J.~Steinhardt,
Phys.\ Rev.\ D \textbf{25}, 2074 (1982). 

\bibitem{kaj}   K.~Kajantie,
  Phys.\ Lett.\ B {\bf 285}, 331 (1992).

\bibitem{m04}  A.~Megevand,
  Phys.\ Rev.\ D {\bf 69}, 103521 (2004)
  [hep-ph/0312305].

\bibitem{hhrw13}    M.~Hindmarsh, S.~J.~Huber, K.~Rummukainen and
    D.~J.~Weir,
  arXiv:1304.2433 [hep-ph].

\bibitem{mssm} M.~S.~Carena, M.~Quiros and C.~E.~M.~Wagner,
Nucl.\ Phys.\ B \textbf{524}, 3 (1998) [arXiv:hep-ph/9710401].
J.~A.~Casas, J.~R.~Espinosa, M.~Quiros and
    A.~Riotto,
Nucl.\ Phys.\ B \textbf{436}, 3 (1995)  [Erratum-ibid.\ B
\textbf{439}, 466
(1995)]  [arXiv:hep-ph/9407389]  
M.~S.~Carena, M.~Quiros and C.~E.~M.~Wagner,
Phys.\ Lett.\ B \textbf{380}, 81 (1996) [arXiv:hep-ph/9603420].
M.~S.~Carena and C.~E.~M.~Wagner,
Nucl.\ Phys.\ B \textbf{452}, 45 (1995) [arXiv:hep-ph/9408253];
J.~R.~Espinosa,
Nucl.\ Phys.\ B \textbf{475}, 273 (1996)  [arXiv:hep-ph/9604320];
J.~E.~Bagnasco and M.~Dine,
Phys.\ Lett.\ B \textbf{303}, 308 (1993)  [arXiv:hep-ph/9212288];
P.~Arnold and O.~Espinosa,
Phys.\ Rev.\ D \textbf{47}, 3546 (1993)  [Erratum-ibid.\ D
\textbf{50}, 6662
(1994)]  [arXiv:hep-ph/9212235];  
Z.~Fodor and A.~Hebecker,
Nucl.\ Phys.\ B \textbf{432}, 127 (1994)  [arXiv:hep-ph/9403219].


\bibitem{scalars} M.~Dine, P.~Huet, R.~L.~Singleton and L.~Susskind,
Phys.\ Lett.\ B \textbf{257}, 351 (1991); 
M.~Dine, P.~Huet and R.~L.~.~Singleton,
Nucl.\ Phys.\ B \textbf{375}, 625 (1992); 
J.~Choi and R.~R.~Volkas,
Phys.\ Lett.\ B \textbf{317}, 385 (1993) [arXiv:hep-ph/9308234];
S.~W.~Ham, Y.~S.~Jeong and S.~K.~Oh,
J.\ Phys.\ G \textbf{31}, 857 (2005) [arXiv:hep-ph/0411352];
J.~R.~Espinosa and M.~Quiros,
Phys.\ Lett.\ B \textbf{305}, 98 (1993) [arXiv:hep-ph/9301285];
A.~Ahriche,
Phys.\ Rev.\ D \textbf{75}, 083522 (2007) [arXiv:hep-ph/0701192];
S.~Profumo, M.~J.~Ramsey-Musolf and G.~Shaughnessy,
JHEP \textbf{0708}, 010 (2007) [arXiv:0705.2425 [hep-ph]];
A.~Ashoorioon and T.~Konstandin,
JHEP {\bf 0907}, 086 (2009) [arXiv:0904.0353 [hep-ph]].
J.~R.~Espinosa and M.~Quiros,
Phys.\ Rev.\ D \textbf{76}, 076004 (2007) [arXiv:hep-ph/0701145].


\bibitem{ms10}
  A.~Megevand and A.~D.~Sanchez,
  Nucl.\ Phys.\ B {\bf 825}, 151 (2010)
  [arXiv:0908.3663 [hep-ph]].


\bibitem{cmqw05} M.~S.~Carena, A.~Megevand, M.~Quiros and
    C.~E.~M.~Wagner,
Nucl.\ Phys.\ B \textbf{716}, 319 (2005) [arXiv:hep-ph/0410352].


\bibitem{ms08}  A.~Megevand and A.~D.~Sanchez,
  Phys.\ Rev.\ D {\bf 77}, 063519 (2008)
  [arXiv:0712.1031 [hep-ph]].

\bibitem{soj97}    G.~Sigl, A.~V.~Olinto and K.~Jedamzik,
  Phys.\ Rev.\ D {\bf 55}, 4582 (1997)
  [astro-ph/9610201].

\bibitem{n11}   J.~M.~No,
  Phys.\ Rev.\ D {\bf 84}, 124025 (2011)
  [arXiv:1103.2159 [hep-ph]].

\bibitem{cn12}   C.~Caprini and J.~M.~No,
  JCAP {\bf 1201}, 031 (2012)
  [arXiv:1111.1726 [hep-ph]].

\bibitem{fa90}   K.~Freese and F.~C.~Adams,
  Phys.\ Rev.\ D {\bf 41}, 2449 (1990).

\bibitem{kf92}
  M.~Kamionkowski and K.~Freese,
  Phys.\ Rev.\ Lett.\  {\bf 69}, 2743 (1992)
  [hep-ph/9208202].

\bibitem{a97}
  M.~Abney,
  Phys.\ Rev.\ D {\bf 55}, 582 (1997)
  [hep-ph/9606476].

\bibitem{m01}   A.~Megevand,
  Phys.\ Rev.\ D {\bf 64}, 027303 (2001)
  [hep-ph/0011019].

\end{thebibliography}
\end{document}